\documentclass[longauth]{aa}
\usepackage{graphicx}
\usepackage[colorlinks=true, linkcolor=blue, citecolor=blue, urlcolor=blue]{hyperref}
\usepackage{txfonts}
\usepackage{lscape}
\usepackage{epstopdf}
\usepackage{CJK}
\usepackage{float}
\usepackage[absolute]{textpos}
\usepackage{amsmath}
\usepackage{mathrsfs}
\usepackage{adjustbox}
\usepackage[T1]{fontenc}
\usepackage{orcidlink}
\usepackage[switch]{lineno}
\usepackage{rotating}
\usepackage{subcaption}

\begin{document}

\begin{CJK*}{UTF8}{gbsn}

\title{The ALMA-ATOMS survey: Vibrationally excited HC$_3$N lines in hot cores}

\author{
Li Chen\inst{1}\orcidlink{0009-0009-8154-4205}\and
Sheng-Li Qin\inst{1}\orcidlink{0000-0003-2302-0613}\and
Tie Liu\inst{2}\orcidlink{0000-0002-5286-2564}\and
Paul F. Goldsmith\inst{3}\orcidlink{0000-0002-6622-8396}\and
Xunchuan Liu\inst{2}\orcidlink{0000-0001-8315-4248}\and
Yaping Peng\inst{4}\orcidlink{0000-0001-5703-1420}\and
Xindi Tang\inst{5,6}\orcidlink{0000-0002-4154-4309}\and
Guido Garay\inst{7,8}\and
Zhiping Kou\inst{5,9,1}\and
Mengyao Tang\inst{10}\orcidlink{0000-0001-9160-2944}\and
Patricio Sanhueza\inst{11,12,13}\orcidlink{0000-0002-7125-7685}\and
Zi-Yang Li\inst{1,2}\orcidlink{0009-0005-7028-0735}\and
Prasanta Gorai\inst{14,15}\orcidlink{0000-0003-1602-6849}\and
Swagat R. Das\inst{7}\orcidlink{0000-0001-7151-0882}\and
Leonardo Bronfman\inst{7}\orcidlink{0000-0002-9574-8454}\and
Lokesh Dewangan\inst{16}\orcidlink{0000-0001-6725-0483}\and
Pablo Garc\'{i}a\inst{17,8}\and
Shanghuo Li\inst{18}\orcidlink{0000-0003-1275-5251}\and
Chang Won Lee\inst{19,20}\orcidlink{0000-0002-3179-6334}\and
Hong-Li Liu\inst{1}\orcidlink{0000-0003-3343-9645}\and
L. Viktor T\'{o}th\inst{21,22}\orcidlink{0000-0002-5310-4212}\and
James O. Chibueze\inst{23,24,25}\orcidlink{0000-0002-9875-7436}\and
Jihye Hwang\inst{19}\orcidlink{0000-0001-7866-2686}\and\\
Xiaohu Li\inst{5,6}\orcidlink{0000-0003-2090-5416}\and
Fengwei Xu\inst{26,27}\orcidlink{0000-0001-5950-1932}\and
Jiahang Zou\inst{1,2}\orcidlink{0009-0000-9090-9960}\and
Wenyu Jiao\inst{2}\orcidlink{0000-0001-9822-7817}\and
Zhenying Zhang\inst{1,2}\orcidlink{0009-0005-4295-5010}\and
Yong Zhang\inst{28}\orcidlink{0000-0002-1086-7922}
}

\institute{
School of Physics and Astronomy, Yunnan University, Kunming 650091, People's Republic of China\\
\email{li.chen@mail.ynu.edu.cn, qin@ynu.edu.cn}
\and
Shanghai Astronomical Observatory, Chinese Academy of Sciences, 80 Nandan Road, Shanghai 200030, People's Republic of China\\
\email{liutie@shao.ac.cn}
\and
Jet Propulsion Laboratory, California Institute of Technology, 4800 Oak Grove Drive, Pasadena CA 91109, USA
\and
Department of Physics, Faculty of Science, Kunming University of Science and Technology, Kunming 650500, People's Republic of China
\and
Xinjiang Astronomical Observatory, Chinese Academy of Sciences,
150 Science 1-Stree, Urumqi, Xinjiang 830011, People's Republic of China
\and
Xinjiang Key Laboratory of Radio Astrophysics, 150 Science 1-Street, Urumqi, Xinjiang 830011, People's Republic of China
\and
Departamento de Astronom\'{i}a, Universidad de Chile, Las Condes, 7591245 Santiago, Chile
\and
Chinese Academy of Sciences South America Center for Astronomy, National Astronomical Observatories, CAS, Beĳing 100101, People's Republic of China
\and
University of Chinese Academy of Sciences, Beijing 100049, People's Republic of China
\and
Institute of Astrophysics, School of Physics and Electronical Science, Chuxiong Normal University, Chuxiong 675000, People's Republic of China
\and
Department of Earth and Planetary Sciences, Institute of Science Tokyo, Meguro, Tokyo, 152-8551, Japan
\and
National Astronomical Observatory of Japan, National Institutes of Natural Sciences, 2-21-1 Osawa, Mitaka, Tokyo 181-8588, Japan
\and
Department of Astronomical Science, SOKENDAI (The Graduate University for Advanced Studies), 2-21-1 Osawa, Mitaka, Tokyo 181-8588, Japan
\and
Rosseland Centre for Solar Physics, University of Oslo, PO Box 1029 Blindern, 0315 Oslo, Norway
\and
Institute of Theoretical Astrophysics, University of Oslo, PO Box 1029 Blindern, 0315 Oslo, Norway
\and
Physical Research Laboratory, Navrangpura, Ahmedabad 380 009, India
\and
Instituto de Astronom\"{i}a, Universidad Cat\'{o}lica del Norte, Antofagasta, Chile
\and
Max Planck Institute for Astronomy, K\"{o}nigstuhl 17, D-69117 Heidelberg, Germany
\and
Korea Astronomy and Space Science Institute, 776 Daedeokdaero, Yuseong-gu, Daejeon 34055, Republic of Korea
\and
University of Science and Technology, Korea (UST), 217 Gajeong-ro, Yuseong-gu, Daejeon 34113, Republic of Korea
\and
Department of Astronomy, E\"{o}tv\"{o}s Lor\'{a}nd University, P\'{a}zm\'{a}ny P\'{e}ter s\'{e}t\'{a}ny 1/A, H-1117, Budapest, Hungary 
\and
University of Debrecen, Faculty of Science and Technology, Egyetemt\'{e}r 1, H-4032 Debrecen, Hungary;
\and
Department of Mathematical Sciences, University of South Africa, Cnr Christian de Wet Rd and Pioneer Avenue, Florida Park, 1709 Roodepoort, South Africa
\and
Centre for Space Research, North-West University, Potchefstroom Campus, Private Bag X6001, Potchefstroom 2520, South Africa
\and
Department of Physics and Astronomy, Faculty of Physical Sciences, University of Nigeria, Carver Building, 1 University Road, Nsukka 410001, Nigeria
\and
I. Physikalisches Institut, Universit{\"a}t zu K{\"o}ln, Z{\"u}lpicher Stra{\ss}e 77, 50937 K{\"o}ln, Germany
\and
Kavli Institute for Astronomy and Astrophysics, Peking University, 5 Yiheyuan Road, Haidian District, Beijing 100871, China
\and
School of Physics and Astronomy, Sun Yat-sen University, 2 Daxue Road, Zhuhai, Guangdong 519082, People's Republic of China
}

\abstract
{Interstellar molecules are excellent tools for studying the physical and chemical environments of massive star-forming regions. In particular, the vibrationally excited HC$_3$N (HC$_3$N*) lines are the key tracers for probing hot cores environments.}
{We present the Atacama Large Millimeter/submillimeter Array (ALMA) 3 mm observations of HC$_3$N* lines in 60 hot cores and investigate how the physical conditions affect the excitation of HC$_3$N* transitions.}
{We used the XCLASS for line identification. Under the assumption of local thermodynamic equilibrium, we derived the rotation temperature and column density of HC$_3$N* transitions in hot cores. Additionally, we calculated the H$_2$ column density and number density, along with the abundance of HC$_3$N* relative to H$_2$, to enable a comparison of the physical properties of hot cores with different numbers of HC$_3$N* states.}
{We have detected HC$_3$N* lines in 52 hot cores, 29 of which show more than one vibrationally excited state. Hot cores with higher gas temperatures have more detections of these vibrationally excited lines. The excitation of HC$_3$N* requires dense environments, and its spatial distribution is affected by the presence of UC H{\sc ii} regions. The observed column density of HC$_3$N* contributes to the number of HC$_3$N* states in hot-core environments.}
{After analyzing the various factors influencing HC$_3$N* excitation in hot cores, we conclude that the excitation of HC$_3$N* is mainly driven by mid-IR pumping, while collisional excitation is ineffective.}


\maketitle

\section{Introduction}
Massive star-forming regions represent some of the most dynamic and intriguing environments in the cosmos. These areas serve as cosmic crucibles where gravitational forces, radiation, and turbulent motion interact to shape the birth and evolution of stellar progenitors \citep{2017MNRAS.469.2286Z, 2018ApJ...856..141T, 2018ARA&A..56...41M, 2020ApJ...894L..14L, 2021RAA....21...14Y, 2022MNRAS.516.1983S, 2022ApJ...941..202R, 2024RAA....24f5003L}. Motivated by the fundamental importance of massive star-forming regions, researchers have long sought to dissect the molecular landscapes that characterize these environments (for comprehensive reviews, see \citealt{2009ARA&A..47..427H, 2016ARA&A..54..181G, 2018IAUS..332....3V, 2020ARA&A..58..727J}). Among the myriad molecular species, cyanoacetylene (HC$_3$N) stands out as a particularly intriguing tracer of dense and warm gas surrounding newly forming stars \citep{1991JKAS...24..217C, 1996ApJ...460..343B, 2019MNRAS.489.4497Y} and energetic processes in highly obscured regions \citep{2021MNRAS.502.3021R}. Previous observations have laid important groundwork in the investigation of HC$_3$N lines in massive star-forming regions \citep{2017MNRAS.466..248L, 2018ApJ...854..133T, 2021ApJS..253....2H}, shedding light on their abundance patterns, spatial distributions, and excitation mechanisms.

The shortest cyanopolyyne, HC$_3$N (H--C$\equiv$C--C$\equiv$N), possesses one CN triple bond and one CC triple bond. As a linear molecule, HC$_3$N is widespread in the interstellar medium (ISM) and has been detected in various astronomical sources both in the Milky Way \citep{1971ApJ...163L..35T, 1981Natur.292..686K, 2012ApJ...756...12L, 2015A&A...575A..84V, 2021MNRAS.506L..45Z, 2024ApJS..271....3L} and in external galaxies \citep{1990A&A...236...63M, 2020ApJ...889..129W, 2021A&A...656A..46M}. The excitation of HC$_3$N can be driven by the absorption of mid-IR photons and/or collisions with H$_2$ \citep{2020MNRAS.491.4573R}. Because the linear structure of HC$_3$N lacks a center of symmetry, the spectra produced by its transitions are complex. Experiments by \cite{1986JMoSp.116..384Y} and \cite{2000JMoSp.204..133T} have given information on the transitions of rotational spectra of various vibrationally excited HC$_3$N states in the millimeter and submillimeter wave bands. HC$_3$N exhibits seven normal vibrations: four stretching vibrations (v$_1$-v$_4$) of $\Sigma$ symmetry, and three doubly degenerate bending vibrations (v$_5$-v$_7$) of $\Pi$ symmetry \citep{1976MolPh..32..473M}. The characteristics of these seven vibrational states are summarized in Table~\ref{tab:mdoes} \citep{2018PhRvA..98e2708R}. The vibrationally excited state of HC$_3$N refers to a specific state with higher vibrational energy, resulting from the excitation of a vibrational mode after the molecule absorbs energy. Experiments show that higher vibrational energy levels require a higher impact energy to be collisionally excited 
\citep{2014JChPh.140q4305L}. In astrophysical environments, the critical densities (n$\rm _{crit}$) for collisional excitation of the v$_7$=1, v$_6$=1, v$_5$=1, and v$_4$=1 levels at 300 K are 4$\times$10$^8$, 3$\times$10$^{11}$, 7$\times$10$^{12}$, and 2$\times$10$^{10}$ cm$^{-3}$, respectively \citep{1999A&A...341..882W}. Therefore, in this work, only the low-lying vibrational modes (v$_4$, v$_5$, v$_6$, and v$_7$) with one or multiple vibrationally excited states are investigated, as detected in the ALMA band 3 survey.

Despite significant progress, gaps persist in our understanding, particularly in relation to the precise role of vibrationally excited HC$_3$N (abbreviated as HC$_3$N*) in probing the conditions prevailing within hot cores. Clarifying these issues is essential for constructing comprehensive models of massive star formation and refining our understanding of the underlying physics. Spurred by the potential to unveil crucial details about the physical conditions operating in extreme environments, attention has turned to the study of HC$_3$N* lines within hot cores \citep{1982ApJ...260..147G, 1999A&A...341..882W, 2000A&A...361.1058D, 2010A&A...515A..71C, 2013A&A...559A..47B, 2017ApJ...837...49P, 2017A&A...604A..32P, 2022ApJ...931...99T}. The spectral information of the vibrationally excited states can reveal the behavior of HC$_3$N molecules in different energy states and their distribution in the ISM. Particularly in high-temperature environments, the study of vibrationally excited states is significant for understanding molecular excitation and chemical reaction processes \citep{2013A&A...559A..51E}.

Against this backdrop, the primary objective of this paper is to present a detailed examination of HC$_3$N* lines in massive star-forming regions and provide a comprehensive understanding of the role HC$_3$N* plays in studying the inner physical conditions of hot cores. By systematically exploring the characteristics of different HC$_3$N* states, this study intends to elucidate their observational properties and the physical processes shaping their spectral lines. The structure of this paper is organized as follows: In Sect. \ref{obs} we describe the ALMA observations and the data reduction process. Section \ref{met} presents the fundamental results on the detection of HC$_3$N* lines and the preliminary analysis of the derived physical parameters. In Sect. \ref{dis} we discuss the results of the study in depth, identifying the physical properties and the vibrational excitation conditions of HC$_3$N. Finally, Sect. \ref{con} outlines the key conclusions of this paper.

\begin{table}
\centering
\caption{Lowest vibrational modes of HC$_3$N}
\label{tab:mdoes}
\begin{tabular}{cccc}
\hline\hline
Mode & Vibration & \multicolumn{2}{c}{Energy$^{a}$} \\
\cline{3-4}
 & & (meV) & (K)\\
\hline
v$_1$ & C--H stretch & 412 & 4781 \\
v$_2$ & C$\equiv$N stretch & 282 & 3272 \\
v$_3$ & C$\equiv$C stretch & 257 & 2982 \\
v$_4$ & C--C stretch & 109 & 1265 \\
v$_5$ & CCH bend & 82 & 952 \\
v$_6$ & CCN bend & 62 & 719 \\
v$_7$ & CCC bend & 28 & 325 \\
\hline
\end{tabular}
\begin{flushleft}
{$^{a.}${} The spectroscopic experimental vibrational energies are from \cite{2014JChPh.140q4305L}}
\end{flushleft}
\end{table}

\section{Observations and data reduction} \label{obs}

\subsection{ALMA observations}

The ATOMS, standing for ALMA Three-millimeter Observations of Massive Star-forming regions, is an ALMA band 3 survey aimed at studying the physical and chemical conditions of 146 massive star-forming regions (\citealt{2020MNRAS.496.2790L}; Project ID: 2019.1.00685.S; PI: Tie Liu). The observations were conducted from September to mid-November 2019 and included both the Atacama Compact 7 m Array (ACA) and the 12 m array (C43-2 or C43-3 configuration). SPWs 7 and 8 are tuned in the upper sideband of the observations, which have a bandwidth of 1875 MHz and a spectral resolution of approximately 1.6 km s$^{-1}$. SPWs 7 and 8 were primarily used for continuum imaging and line surveys, with center frequencies of $\sim$ 98,505 MHz and 100,454 MHz, respectively. In the ATOMS project, due to the high resolution and minimal flux loss in the dense molecular line tracers, we exclusively rely on 12 m array data to identify molecular lines and hot cores.

\subsection{Data calibration and imaging}

Data reduction and image processing were conducted using version 5.6 of the Common Astronomy Software Applications \citep[CASA;][]{2007ASPC..376..127M, 2022PASP..134k4501C}. The ACA data and 12 m array data were calibrated separately, and the 12 m array data were chosen for identifying cloud cores and molecular lines. For the continuum maps, line-free channels were selected in SPW 7 and SPW 8, with a central frequency of approximately 99,404 MHz. Line cubes were produced for each SPW with their native spectral resolution, and the maximum recoverable angular scales were approximately 60 arcsec. All images were corrected for primary beam. The 12 m data for the 146 clumps have an angular resolution of approximately 1.2-1.9 arcsec (the linear scale angular resolution are listed in Col. 4 of Tables~\ref{tab:param1} to~\ref{tab:param4}), with a mean sensitivity of approximately 0.4 mJy beam$^{-1}$ for the continuum and better than 10 mJy beam$^{-1}$ per channel for the lines. By using three complex organic molecules (COMs), namely C$_2$H$_5$CN, CH$_3$OCHO, and CH$_3$OH, as probes for hot cores, \cite{2022MNRAS.511.3463Q} successfully identified 60 hot cores, which were selected as the sample for our study.

\section{Methods and results} \label{met}

\subsection{HC$_3$N* line identification}

\begin{figure*}
\centering
\includegraphics[width=17cm,height=4.9cm]{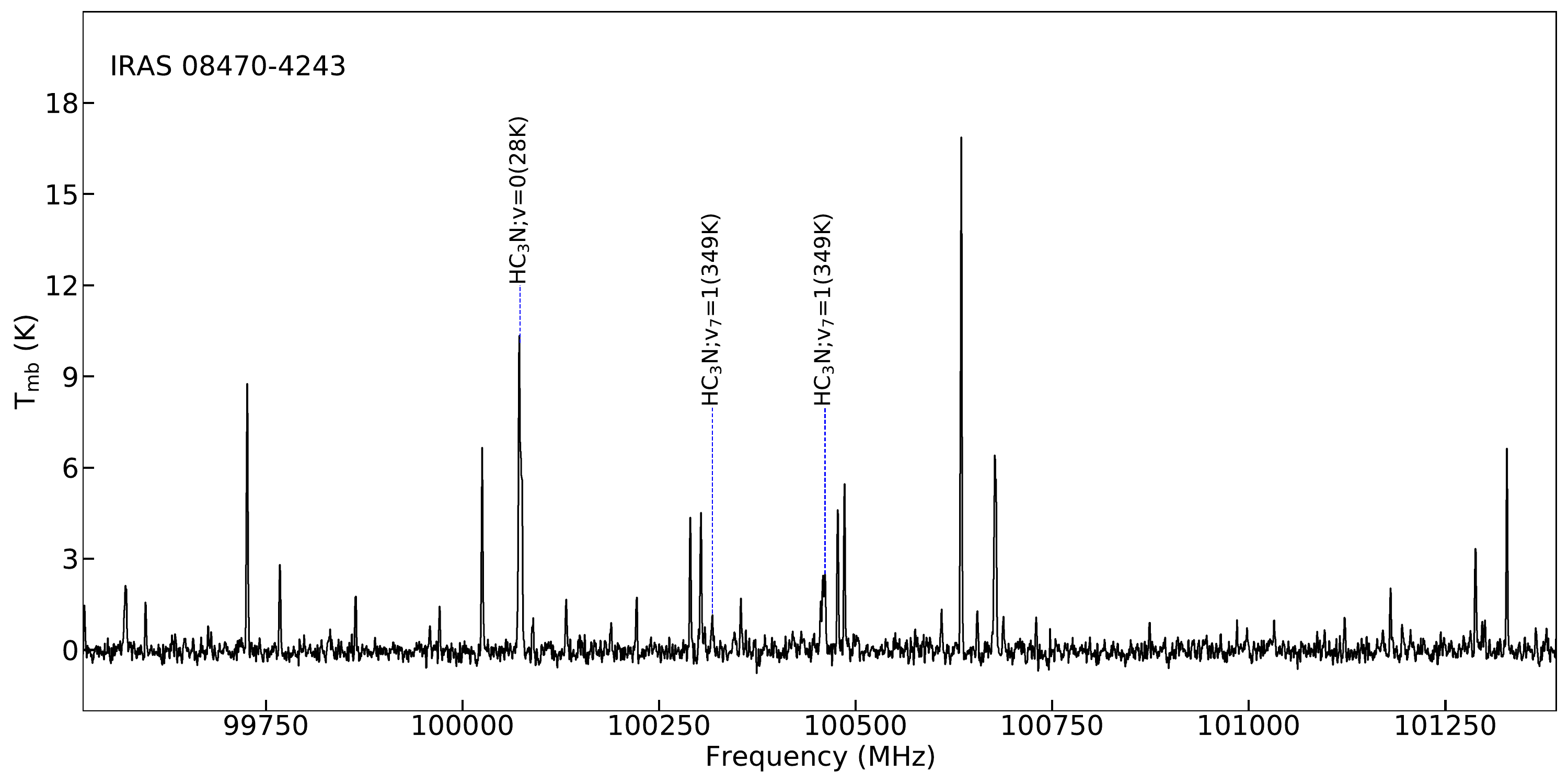}
\includegraphics[width=17cm,height=4.9cm]{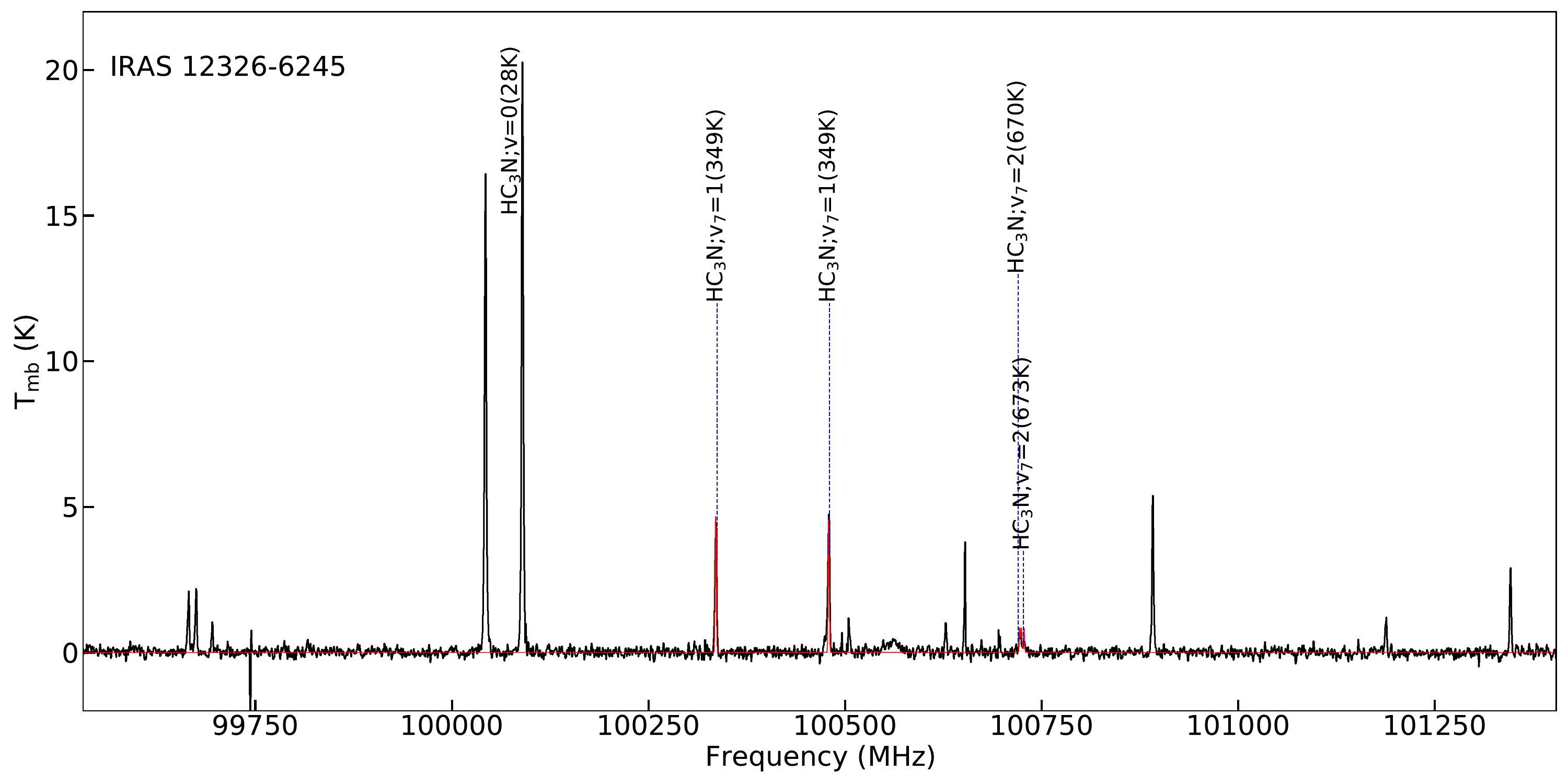}
\includegraphics[width=17cm,height=4.9cm]{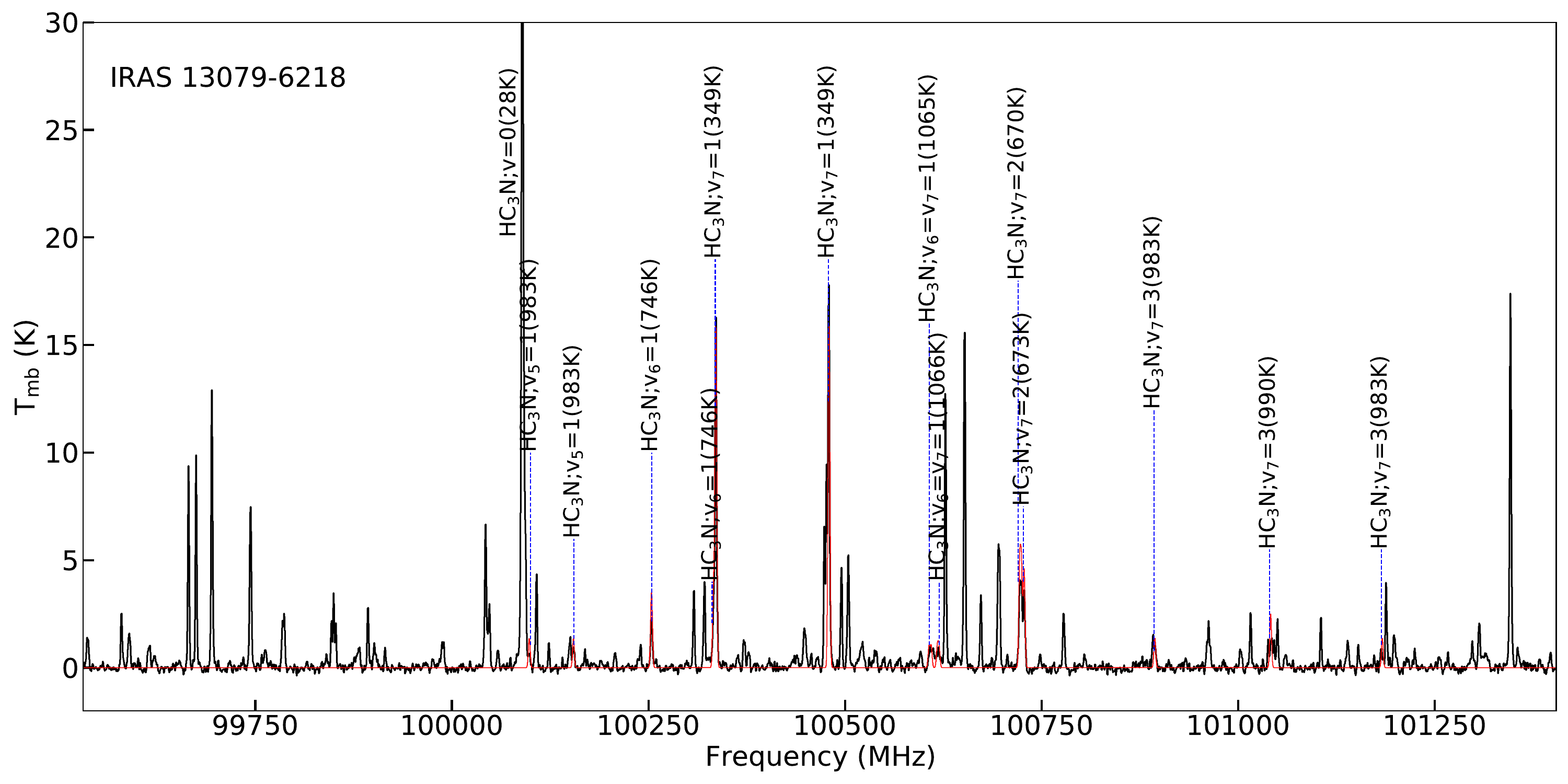}
\includegraphics[width=17cm,height=4.9cm]{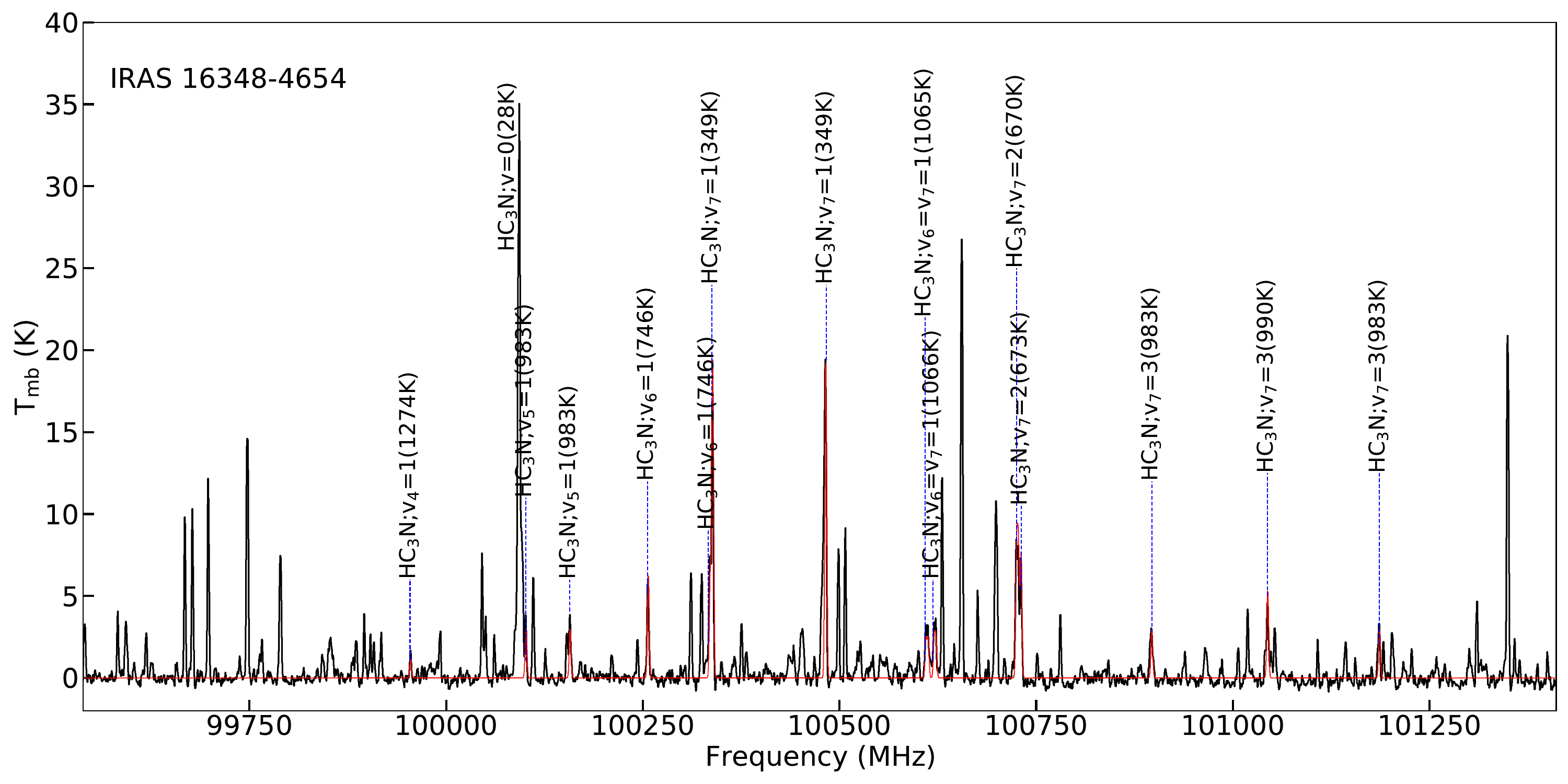}
\caption{Sample spectra of HC$_3$N* in SPW 8 for four typical hot cores. The black lines show the observed spectra at sky frequencies, and the red lines show the XCLASS modeled spectra using the best-fit parameters for vibrationally excited HC$_3$N lines. The complete spectra in SPW 8 for other hot cores are available in Appendix D (on \href{https://doi.org/10.5281/zenodo.14557138}{Zenodo}).\label{fig:spec}}
\end{figure*}

We conducted a search for all vibrationally excited lines of HC$_3$N in 60 hot cores. The molecular lines were obtained from the peak position of each hot core. Precise line identification was performed using the eXtended CASA Line Analysis Software Suite (XCLASS; \citealt{2017A&A...598A...7M})\footnote{\url{https://xclass.astro.uni-koeln.de/}}, assuming local thermodynamic equilibrium (LTE). The Cologne Database for Molecular Spectroscopy (CDMS; \citealt{2001A&A...370L..49M, 2005JMoSt.742..215M, 2016JMoSp.327...95E})\footnote{\url{https://cdms.astro.uni-koeln.de/}} and the Jet Propulsion Laboratory (JPL) molecular databases \citep{1998JQSRT..60..883P}\footnote{\url{http://spec.jpl.nasa.gov}}, accessed through XCLASS, provided line parameters such as rest frequency, quantum number (QN), dipole-weighted transition dipole matrix elements (S$\rm_{ij}\mu^2$), Einstein A coefficient (A$\rm_{ij}$), and upper energy (E$_{u}$). In this survey, we detected a total of seven types of HC$_3$N* lines in 52 hot cores: v$_7$=1, v$_7$=2, v$_7$=3, v$_6$=1, v$_6$=v$_7$=1, v$_5$=1, and v$_4$=1. The seven HC$_3$N states include 18 transitions in SPW 8 and none in SPW 7, all corresponding to the same J=11$-$10 transition. The relevant parameters are listed in Table~\ref{tab:HC3N}. One HC$_3$N v=0 (J=11$-$10) line was also detected in SPW 8 at the rest frequency $\sim$ 100,076 MHz, which indicates that all 52 hot cores contain HC$_3$N molecules. The determination of each line was based on its velocity offset, which was almost the same as the systemic source velocity (see \cite{2020MNRAS.496.2790L} for specific values), and with a line intensity above the 3$\sigma$ detection level.

The distribution of HC$_3$N* lines is concentrated in SPW 8, with a broad range of upper energy levels ranging from 349 to 1274 K. We used XCLASS to fit the physical parameters, including rotation temperature, column density, line width, and velocity offset, by setting the deconvolved size of the continuum core, rotation temperature, source-averaged column density, line width, and velocity offset as free parameters in the fitting process, with a setup of an isotope ratio of 1 (i.e., the same abundance) for HC$_3$N* in each vibrationally excited state. The physical parameters for each vibrationally excited state were determined simultaneously only when there were three or more corresponding lines. However, for hot cores with only v$_7$=1 lines, it was not possible to derive the physical parameters as only two emission lines were detected. Ultimately, we were able to fit 29 hot cores with more than three vibrationally excited lines. The catalog of detected HC$_3$N* lines toward hot cores, along with the corresponding rotation temperatures and column densities, is presented in Table \ref{tab:catalog}. Eight of the 60 hot cores in our sample show no HC$_3$N* emission. This accounts for 13\% of the total detections. Additionally, 23 hot cores exclusively display v$_7$ = 1 emission, which represents 38\% of the total hot core detections. The detection rates for v$_7$=1, v$_7$=2, v$_6$=1, v$_5$=1, v$_7$=3, v$_6$=v$_7$=1, and v$_4$=1 are 87\%, 48\%, 45\%, 27\%, 27\%, 25\%, and 3\%, respectively. 
In Fig.~\ref{fig:spec} we present sample spectra of HC$_3$N toward four representative hot cores, each with varying numbers of HC$_3$N* states, along with the modeled spectra based on the best-fitting parameters. The differences in excitation of different vibrationally excited lines may be attributed to variations in the physical environments of the sources. 

\subsection{Core parameters}

\subsubsection{Continuum flux}

Ultra-compact (UC) H{\sc ii} regions are manifestations of newly formed massive stars that are still embedded in their natal molecular clouds \citep{2002ARA&A..40...27C}. For hot cores without UC H{\sc ii} regions, the continuum flux is mainly contributed by dust emission. We used the 2D Gaussian fitting in CASA to determine the deconvolved core size, where the full width at half maximum (FWHM) along the major and minor axes is denoted as $\rm \theta_{maj}$ and $\rm \theta_{min}$, respectively. An effective core radius R$\rm _{core}$ was calculated using R$\rm _{core}$=$\sqrt{\theta_{\rm bmaj} \times \theta_{\rm bmin}}$. Simultaneously, we obtained the integrated flux S$\rm_\nu^{int}$ and the peak flux S$\rm_\nu^{peak}$ of the hot core. The parameters R$\rm _{core}$, S$\rm_\nu^{int}$, and S$\rm_\nu^{peak}$ for hot cores without UC H{\sc ii} regions are listed in Cols. 5, 7, and 8 of Tables~\ref{tab:param1} and ~\ref{tab:param2}.

For hot cores associated with UC H{\sc ii} regions, we need to subtract the flux of free-free emission of the UC H{\sc ii} regions from the observed continuum maps to obtain the corrected flux of dust emission. The UC H{\sc ii} regions were identified using the H40$\alpha$ line at 99,023 MHz from SPW 7 \citep{2021MNRAS.505.2801L, 2022MNRAS.511.3463Q}. We assumed that the free-free continuum is optically thin at 99,023 MHz. Thus, we can estimate the intensity of free-free continuum emission under the assumption of LTE via \citep{2016era..book.....C}
\begin{equation}
\label{eq:line_ff}
\rm S_{ff} = 1.43 \times 10^{-3} \: \nu^{-1.1} \: {T_e}^{1.15} \: [1 + \frac{N(He^+)}{N(H^+)}] \: \textstyle \int{S_{H40\alpha} d\nu},
\end{equation}
where S$\rm _{ff}$ is the intensity of free-free continuum emission, $\nu$ = 99.023 GHz near the center frequency of the continuum map , T$\rm _e$ is the electron temperature which is assumed as 6000K \citep{1970ApL.....5..167S, 1996ApJS..106..423A, 2022A&A...664A.140K}, N(He$^+$)/N(H$^+$) $\approx$ 0.08 is the typical ratio, and $\rm \int{S_{H40\alpha} d\nu}$ is the integrated intensity of the H40$\alpha$ line. Since molecular gas emission is invariably associated with the presence of dust, this method might slightly overestimate the free-free contribution to the emission from these sources \citep{2024A&A...687A.163B}.

We used the observed continuum emission to subtract the free-free continuum emission and obtain the true flux from the dust,
\begin{equation}
 \label{eq:core_flux}
\rm S_{dust} = S_{obs} - S_{ff}.
\end{equation}
The parameters R$\rm _{core}$, S$\rm_\nu^{int}$, and S$\rm_\nu^{peak}$ for hot cores associated with UC H{\sc ii} regions are listed in Cols. 5, 7, and 8 of Tables~\ref{tab:param3} and ~\ref{tab:param4}. After the correction for dust flux, we found that I18032--2032 core1 and core 4 show almost no dust emission, indicating that the continuum flux is mainly dominated by free-free emission.

\subsubsection{Parameter calculation}

The mass of each core can be calculated as follows \citep{1983QJRAS..24..267H}:
\begin{equation}
 \label{eq:core_mass}
\rm M_{\text{core}} = \frac{D^2 S_\nu^{int} \eta}{\kappa_\nu B_\nu (T_d)},
\end{equation}
where D is the distance to the source, $S\rm _\nu^{int}$ represents the integrated flux of the dust core, $\eta$ is the gas-to-dust ratio, which increases with galactocentric distance R$\rm_{GC}$ , which was described by \citep{2017A&A...606L..12G, 2023ApJ...950...57T},
\begin{equation}
 \label{eq:eta}
\rm \log(\eta) = \left(0.087\pm0.007\right) R_{\text{GC}} + \left(1.44\pm0.03\right),
\end{equation}
where the R$\rm_{GC}$ of the cores was taken from \cite{2020MNRAS.496.2790L}, the dust absorption coefficient $\kappa_\nu$ for molecular cloud cores was interpolated to be 0.24 cm$^2$ g$^{-1}$ at 99,000 MHz \citep{1994A&A...291..943O}, and $\rm B_\nu (T_d)$ is the Planck function at the dust temperature $\rm T_d$ (see footnote a of Table~\ref{tab:param1} for the selection criteria of T$\rm _d$).

From the 3 mm continuum maps, the source-averaged column density of H$_2$ ($\rm N_{H_2}$) can be derived as \citep{2010ApJ...723.1665F, 2019A&A...628A..27B}
\begin{equation}
\label{eq:h2_cd}
\rm N_{\text{$\rm H_2$}} = \frac{S_\nu^{int} \eta}{\mu m_H \Omega  \kappa_\nu B_\nu (T_d)},
\end{equation}
where $\mu\approx2.8$ is the mean particle weight per $\rm H_2$ molecule \citep{2008A&A...487..993K}, $\rm m_H$ is the mass of a hydrogen atom, and $\Omega$ is the solid angle covered by the source.

After the determination of M$\rm_{core}$, the number density of H$_2$, n($\rm{H_2}$), can be derived by
\begin{equation} \label{eq:number_density}
\rm n({H_2}) = \frac{M_{\rm core}}{(4/3) \pi\mu m_{\rm H}R_{\rm core}^3}.
\end{equation}

Tables~\ref{tab:param1} to \ref{tab:param4} list the core mass, H$_2$ column density, H$_2$ number density, and abundance of HC$_3$N* with respect to H$_2$. The source-averaged column density of HC$_3$N* ranges from (6.9$\pm$0.3)$\times$10$^{15}$ to (1.7$\pm$0.1)$\times$10$^{18}$ cm$^{-2}$. It is noteworthy that the HC$_3$N* column density of I18056-1952 is at least one order of magnitude higher than for other hot cores, while the average column density of the remaining hot cores is (2.7$\pm$0.1)$\times$10$^{16}$ cm$^{-2}$. There are no significant differences in column density between hot cores associated with and without UC H{\sc ii} regions. The calculated N$\rm _{H_2}$ of the 52 hot cores containing HC$_3$N* ranges from (5.3$\pm$0.2)$\times$10$^{22}$ to (2.0$\pm$0.2)$\times$10$^{25}$ cm$^{-2}$, while that of the 8 hot cores that do not contain HC$_3$N* ranges from (8.5$\pm$0.5)$\times$10$^{22}$ to (3.8$\pm$0.5)$\times$10$^{24}$ cm$^{-2}$. For the 29 hot cores we fit, the abundance $\rm f_{HC_3N^*}$ ranges from (6.9$\pm$1.5)$\times$10$^{-10}$ to (2.3$\pm$0.2)$\times$10$^{-7}$. The n(H$_2$) of hot cores containing HC$_3$N* ranges from (1.3$\pm$0.2)$\times$10${^5}$ to (5.0$\pm$0.5)$\times$10$^{8}$ cm$^{-3}$, with a mean value of 3.2$\times$10$^{7}$ cm$^{-3}$, while it ranges from (4.9$\pm$0.4)$\times$10${^5}$ to (4.9$\pm$0.2)$\times$10$^{7}$ cm$^{-3}$, with a mean value of 1.4$\times$10$^{7}$ cm$^{-3}$ for hot cores without HC$_3$N*. The regions containing HC$_3$N* have a relatively high H$_2$ number density.

In addition to using XCLASS for the parameter fitting, we also conducted a rotational temperature diagram (RTD) analysis to determine the rotation temperature of the observed HC$_3$N* lines \citep{1999ApJ...517..209G}. Figure E.1 (on \href{https://zenodo.org/doi/10.5281/zenodo.14557138}{Zenodo)} shows the RTD for HC$_3$N* lines across 29 hot cores, all of which were also analyzed with XCLASS. The rotation temperatures obtained via RTD are consistent with those derived from XCLASS within the errors. This supports the reliability of the rotation temperatures we fit.

\begin{figure}
\centering
\includegraphics[width=3.4 in,  height=3.0 in]{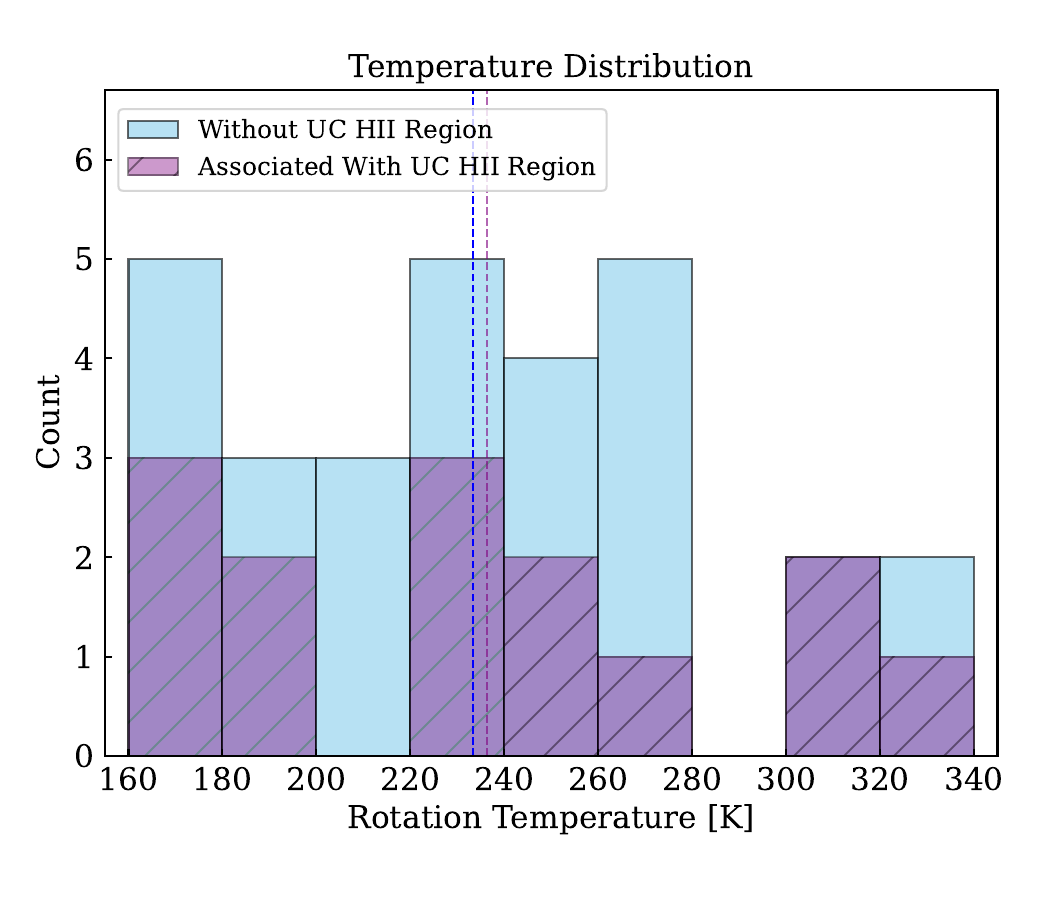}
\caption{Histogram of rotation temperatures of vibrationally excited HC$_3$N lines observed in hot cores. The purple and sky-blue bars present the number of cores associated with and without UC H{\sc ii} regions. The dashed blue line indicates the average rotation temperature of 233 K for hot cores without UC H{\sc ii} regions, and the dashed purple line indicates the average rotation temperature of 236 K for hot cores associated with UC H{\sc ii} regions. 
\label{fig:T}}
\end{figure}

Figure~\ref{fig:T} illustrates the rotation temperature distribution of 29 fit hot cores derived from the XCLASS fit. In our samples, the rotation temperature ranges from (160$\pm$18) to (335$\pm$22) K, with a mean value of (235$\pm$21) K. The median temperature of (235$\pm$48) K is the same as the mean value, indicating that the rotation temperature distribution may be close to symmetric. Furthermore, the mean rotation temperatures of HC$_3$N* molecule in hot cores associated with and without UC H{\sc ii} regions are (236$\pm$17) K and (233$\pm$25) K, respectively, showing no significant difference. This suggests that the presence of UC H{\sc ii} regions likely does not affect the rotation temperature of the HC$_3$N* molecule. 

\begin{figure*}
\centering
\includegraphics[width=\textwidth]{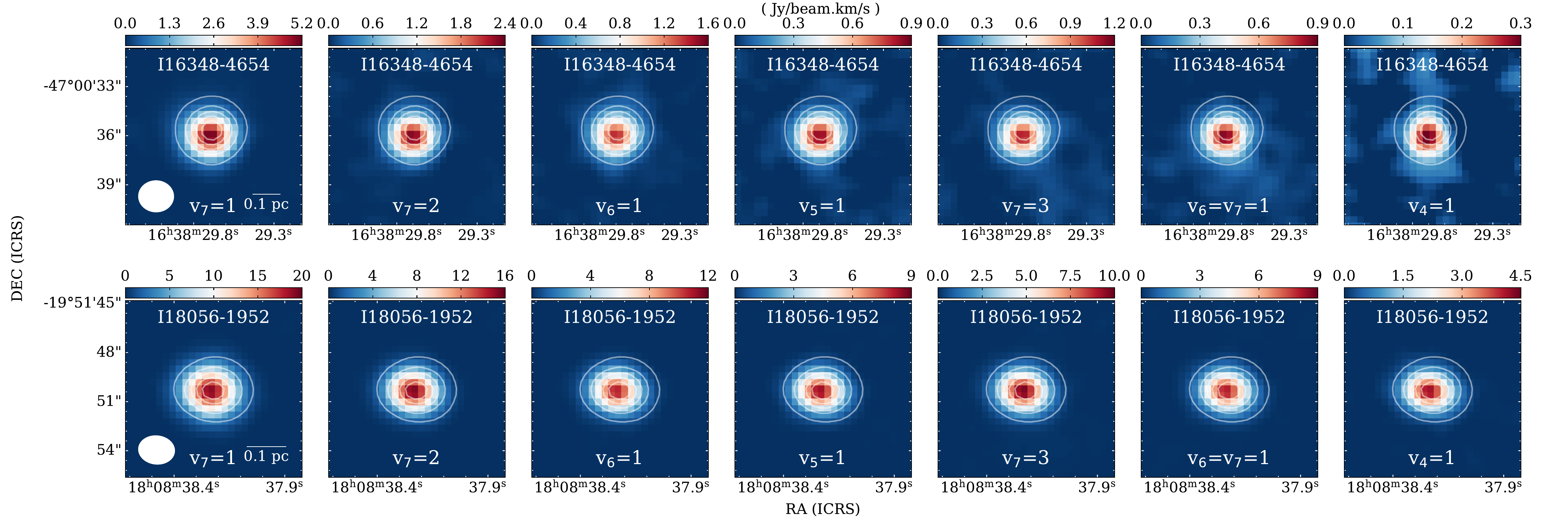}
\caption{Spatial distribution of 7 HC$_3$N* lines of hot cores in IRAS 16348-4654 (top row) and IRAS 18056-1952 (bottom row). The background image of each panel from left to right is  the moment-0 map of the (a) v$_7$=1, $\rm E_u$=349.77530 K, (b) v$_7$=2, $\rm E_u$=670.68918 \& 673.96249 \& 673.96308 K, (c) v$_6$=1, $\rm E_u$=746.53960 K, $\rm E_u$=1065.29153 \& 1066.4191 K, (d) v$_5$=1 $\rm E_u$=983.03920 K, (e) v$_7$=3, $\rm E_u$=990.39731 K, (f) v$_6$=v$_7$=1, and (g) v$_4$=1, $\rm E_u$=1274.35207 K emission. The white contours represent the intensity of the continuum emission, with contour levels ranging from 10\% to 90\% of the peak values in steps of 20\%. The corresponding beam sizes are shown in the bottom left corner, and the linear scales are shown in the bottom right corner of the first image of each panel.
\label{fig:moment0}}
\end{figure*}

\begin{figure*}[htp!]
\centering
\includegraphics[width=7 in,  height=5.4 in]{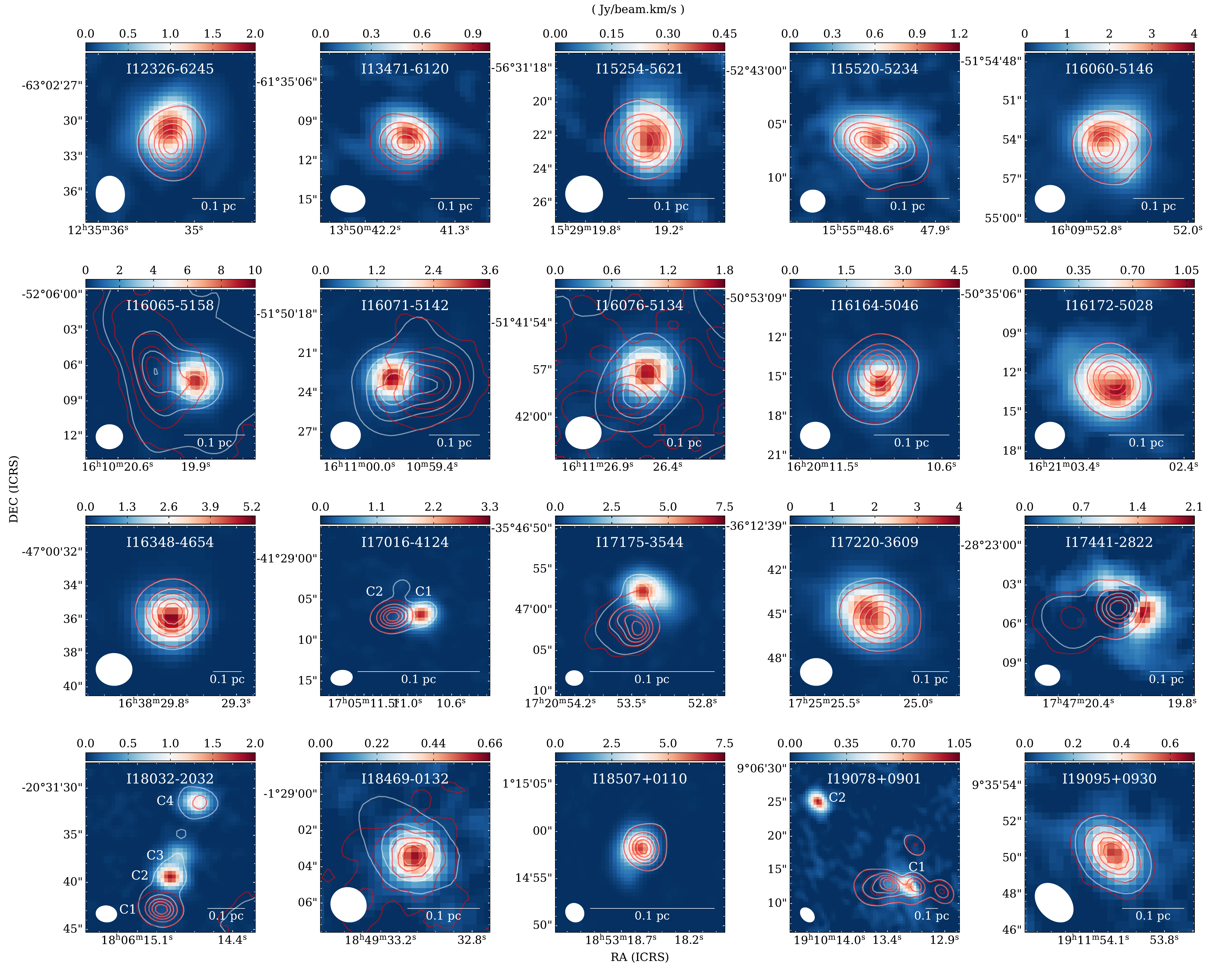}
\caption{Spatial distributions of HC$_3$N v$_7$=1 emission in 20 sources associated with UC H{\sc ii} regions. The background images are moment-0 maps of v$_7$=1 emission. 
The white contours represent the intensity of continuum emission, with contour levels ranging from 10\% to 90\% of the peak values in steps of 20\%. The red contours represent the intensity of UC H{\sc ii} regions (traced by H40$\alpha$ emission), also with contour levels ranging from 10\% to 90\% of the peak values in steps of 20\%. The corresponding beam sizes are shown in the bottom left corner, and the linear scales are shown in the bottom right corner of the image of each source.}
\label{fig:Hii_m0}
\end{figure*}

\begin{figure*}[htp!]
\centering
\includegraphics[width=7 in,  height=6.8 in]{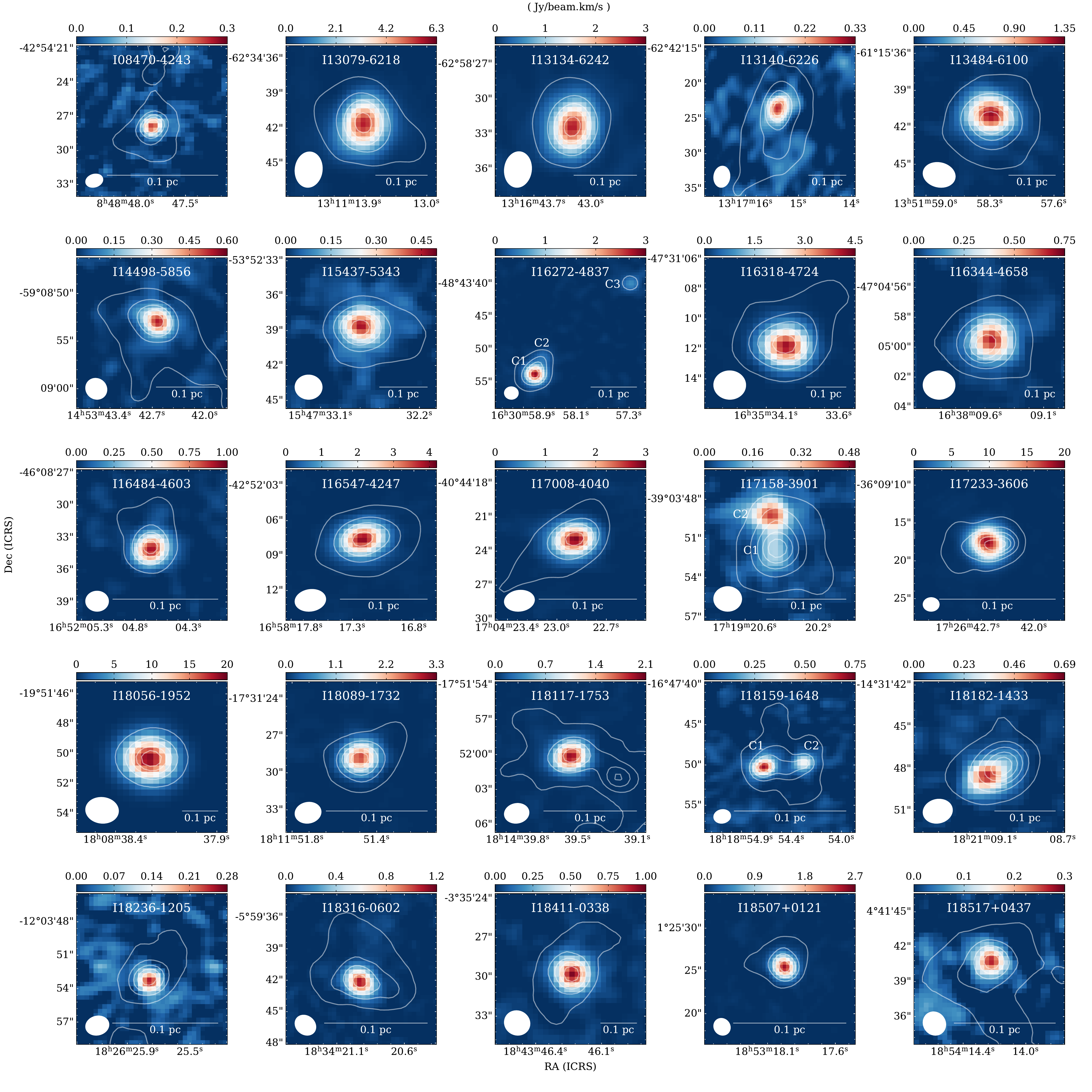}
\caption{Spatial distribution of HC$_3$N v$_7$=1 in 25 sources without UC H{\sc ii} regions. The background images are moment-0 maps of v$_7$=1 emissions. The white contours represent the intensity of continuum emission, with contour levels ranging from 10\% to 90\% of the peak values in steps of 20\%. The corresponding beam sizes are shown in the bottom left corner, and the linear scales are shown in the bottom right corner of the image of each source. \label{fig:no_Hii_m0}}
\end{figure*}

\subsection{Spatial distributions}

Figure~\ref{fig:moment0} shows the example moment-0 maps of 7 HC$_3$N* states of hot cores in IRAS 16348--4654 and IRAS 18056--1952, two sources that detected the seven types of emission, for the analysis of spatial distributions of HC$_3$N* molecules in hot cores. Because these lines selected to create moment-0 maps are not contaminated by other molecular emissions, the resulting distribution is entirely of the vibrationally exited states. The moment-0 maps tentatively suggest a general similarity in the spatial distribution of various types of HC$_3$N* emission. The emissions from different energy levels all originate from the central region of the molecular core, almost without differences in their spatial distribution. However, since these two sources are located at very large distances, we were unable to resolve the emission distribution with 3 mm data. It is also clearly that higher upper energy levels exhibit lower integrated intensities of vibrational excitation, indicating an inverse relation of the observed excitation intensity and the upper energy level. This also implies that it becomes increasingly difficult to excite higher-energy levels. Data with a higher angular resolution will reveal the actual distribution of these transitions and allow us to infer the physical and chemical properties in much greater detail. This is beyond the scope of this work, however.

Using the H40$\alpha$ line at 99,023 MHz from SPW 7, we identified 20 hot cores that are affected by ionized hydrogen gas emission. As shown in Fig. \ref{fig:Hii_m0}, the HC$_3$N* molecular emission peaks and the continuum cores consistently shift in regions associated with UC H{\sc ii} regions. Six of the 20 hot cores with UC H{\sc ii} regions (I12326-6245, I6060-5146, I16076-5134, I16348-4654, I17175-3544, and I17441-2822) displayed significant separations greater than 4500 au, 11 cases (I13471-6120, I15254-5621, I15520-5234, I16065-5158, I16071-5142, I16164-5046, I16172-5028, I17016-4124, I17220-3609, I18507+0110, and I19078+0901c1) showed relatively small separations smaller than 4500 au, and 3 cases (I18032-2032c4, I18469-0132, and I19095+0930) were aligned with the H40$\alpha$ peak. Figure \ref{fig:no_Hii_m0} illustrates the spatial distribution of HC$_3$N* molecules in hot cores without an UC H{\sc ii} region. Of the hot cores without UC H{\sc ii} regions, 28 sources clearly exhibit consistent alignment, while only 4 sources (I13484-6100, I16344-4658, I17233-3606, and I18182-1433) show minor peak shifts, with separations smaller than 4500 au.

\begin{figure*}[htp!]
\centering
\includegraphics[width=0.49\linewidth]{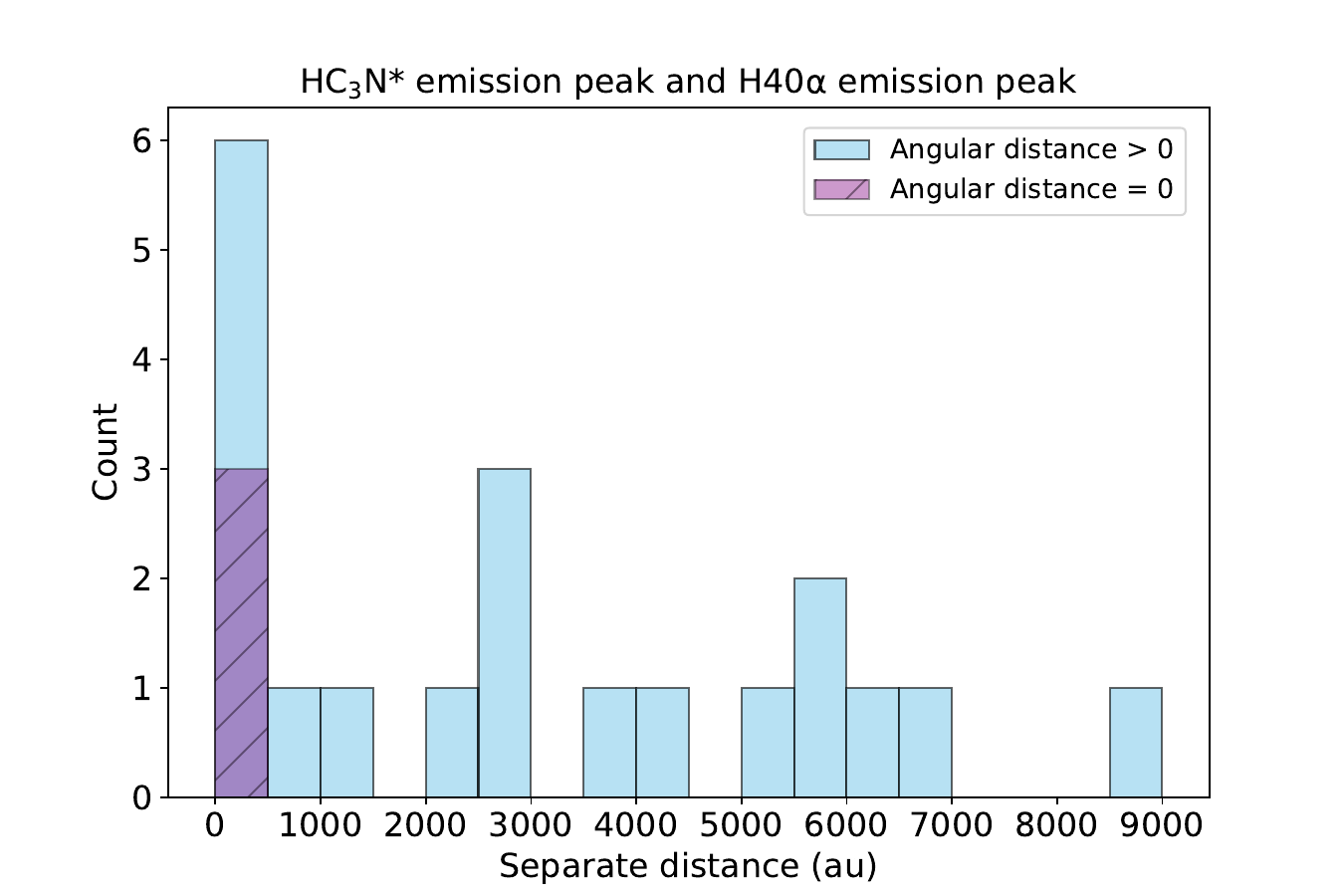}
\includegraphics[width=0.49\linewidth]{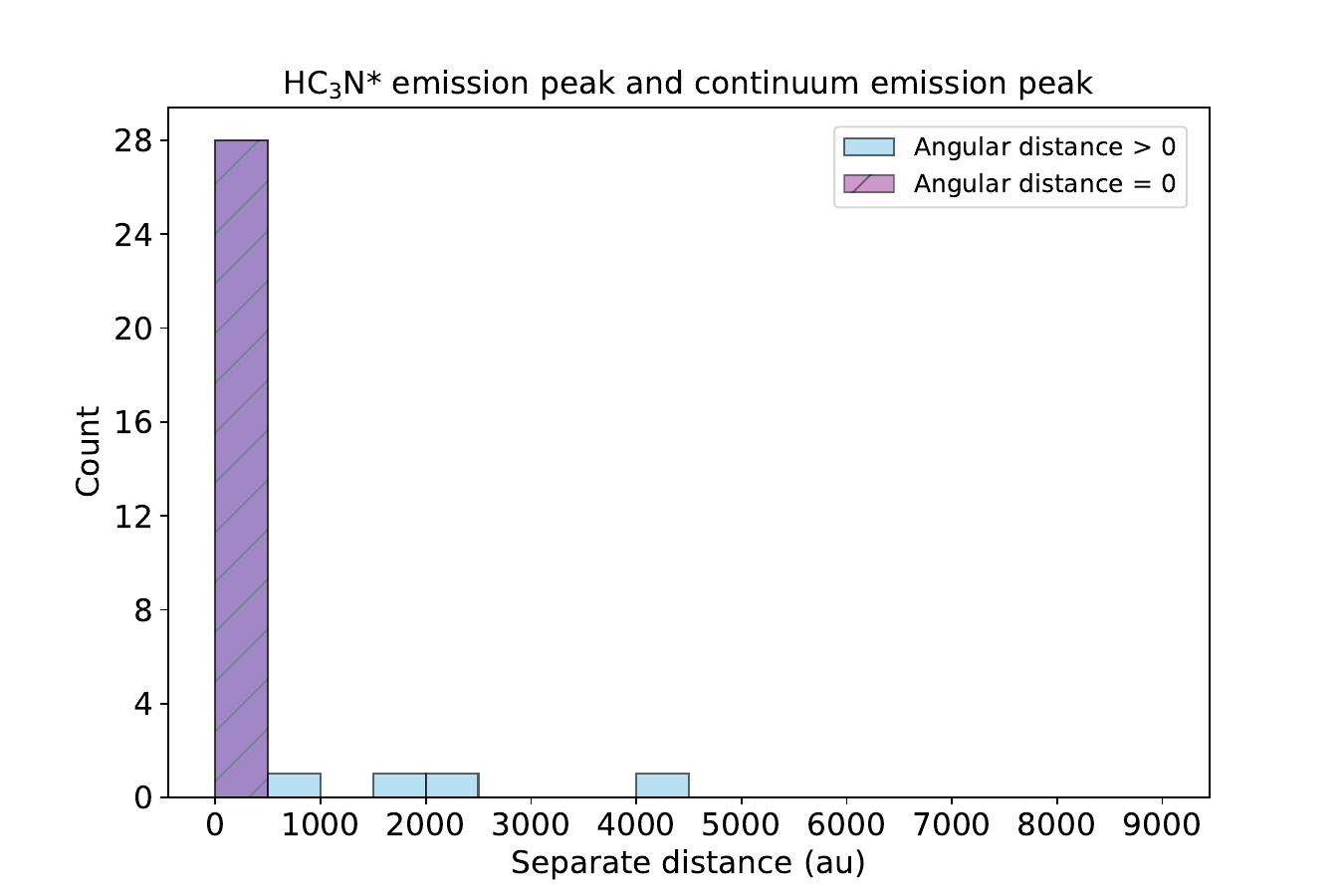}
\caption{Histogram of the separations (angular distances) between the peak of the HC$_3$N* emission and peak of the H40$\alpha$ emission for sources associated with UC H{\sc ii} regions (left panel) and the separations between the peak of the HC$_3$N* emission and peak of the continuum emission for sources without UC H{\sc ii} regions (right panel). The blue bars represent sources with angular distances greater than zero, and the purple bars represent sources with angular distances equal to zero.\label{fig:AD}}
\end{figure*}

\section{Discussion} \label{dis}

\subsection{The role of UC H{\sc ii} regions}
 
The intense ultraviolet (UV) radiation and associated pressure originating from UC H{\sc ii} regions play a crucial role in shaping the motion and structure of the surrounding interstellar medium \citep{2018PhDT........50A}. The left panel of Fig. \ref{fig:AD} provides a histogram of the angular distances between the peaks of HC$_3$N* and H40$\alpha$ emission in sources associated with UC H{\sc ii} regions. In hot cores associated with UC H{\sc ii} regions separated by less than 4500 au, UV radiation has the potential to push interstellar molecules away from the cloud core by radiation pressure. However, at larger separations, the peak shifts may be attributed to two factors. First, photodissociation of HC$_3$N* molecules occurs due to the radiation from UC H{\sc ii} regions. Second, the molecular gas densities at peak positions of UC H{\sc ii} regions fall below the critical densities, and HC$_3$N* cannot be stimulated effectively. These cases suggest that hot cores associated with UC H{\sc ii} regions might be influenced by an energy source that is distinct from the source that drives the UC H{\sc ii} region itself. The right panel of Fig. \ref{fig:AD} displays a histogram of the angular distances between the peaks of HC$_3$N* and continuum emission in sources without UC H{\sc ii} regions. In most hot cores without UC H{\sc ii} regions, where the peaks of HC$3$N* and continuum emissions are well aligned, the heating source is likely stars (or protostars) with luminosities below 10$^3$ L$_{\sun}$, which is insufficient to generate an H{\sc ii} region through UV photon emission. This hypothesis is further supported by the observation that half of the cores listed in Table \ref{tab:param1} have masses below 50 M$_{\sun}$, which is typically inadequate to form luminous protostars. In a few cases, hot cores may be heated by luminous protostars with strong UV radiation, but the appearance of an H{\sc ii} region is quenched by the infalling gas.

\begin{figure*}[htp!]
\centering
\includegraphics[width=3.2 in,  height=2.6 in]{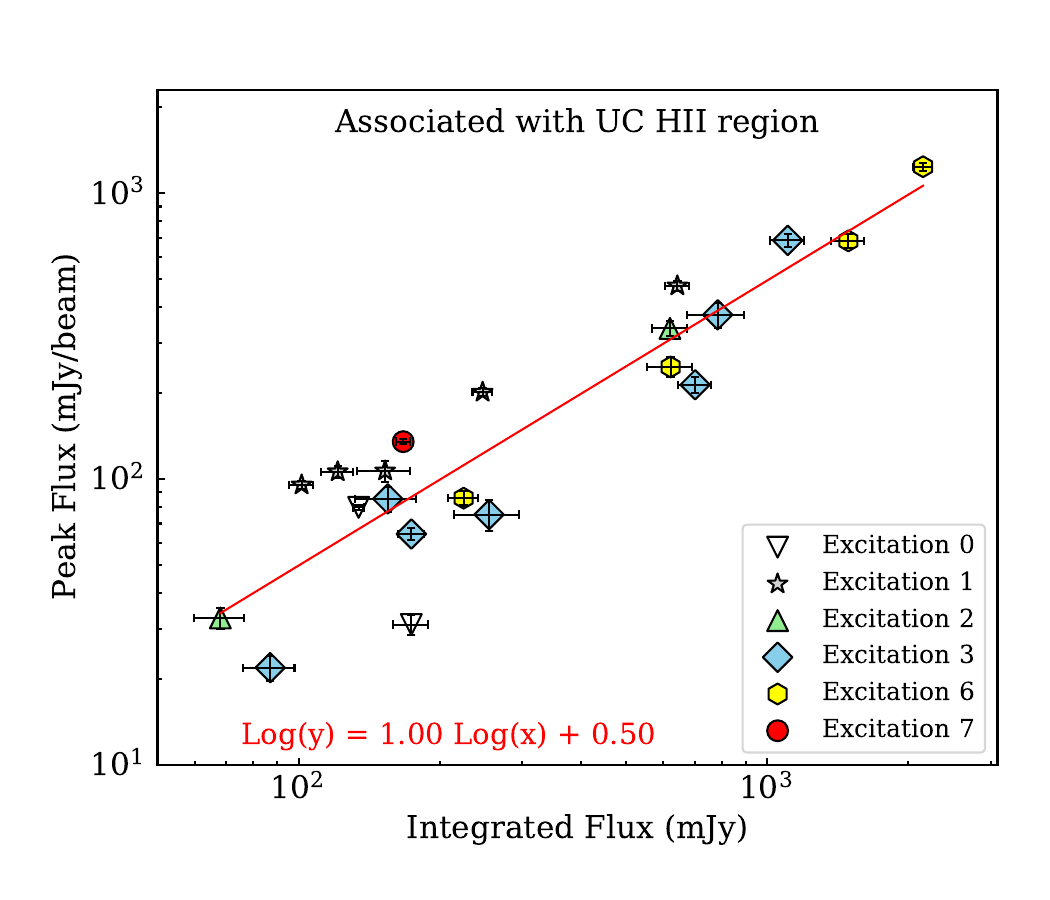}
\includegraphics[width=3.2 in,  height=2.6 in]{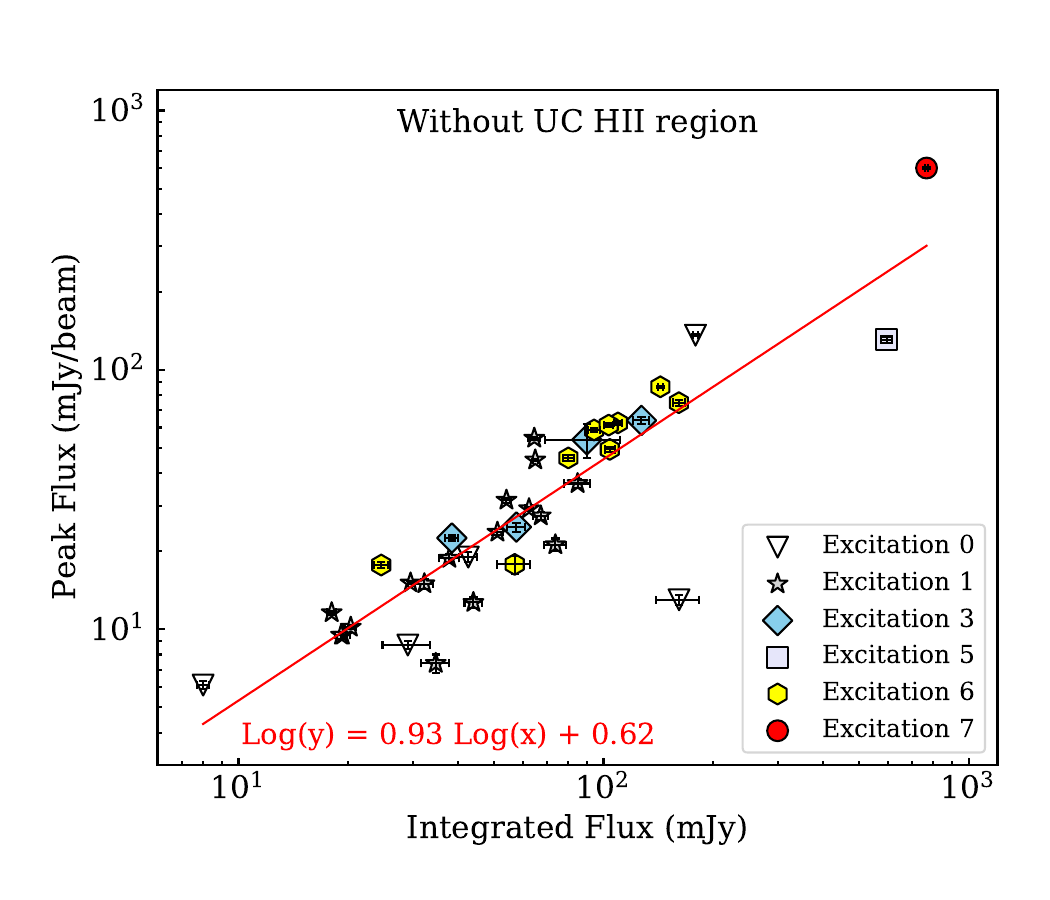}
\caption{Scatter plot of the peak flux vs. integrated flux for sources associated with UC H{\sc ii} regions (left panel) and for sources without UC H{\sc ii} regions (right panel). The different markers represent different types of hot cores that possess different numbers of HC$_3$N* states. The linear least-squares fit for all dots is shown as the solid red line.\label{fig:FF}}
\end{figure*}

We categorized the types of hot cores based on the number of HC$_3$N* states and refer to each category as a different excitation type (e.g., excitation type 2 means that a hot core has HC$_3$N v$_7$=1 and v$_7$=2 lines, and excitation type 3 means that a hot core has HC$_3$N v$_7$=1, v$_7$=2, and v$_6$=1 lines.). Figure~\ref{fig:FF} illustrates the correlation between the integrated flux and the peak flux of dust associated with and without UC H{\sc ii} regions. After the correction for continuum flux by deducting free-free emission, hot cores with higher energy levels of excitation types tend to have higher dust flux. The ratio of the peak intensity to the integrated intensity can indicate whether a source is compact or extended. A higher ratio suggests a more compact source. The hot cores with HC$_3$N* excitation have a strong linear relation of the peak dust flux and the integrated dust flux, indicating that these sources are relatively compact. In contrast, hot cores without HC$_3$N* excitation or with v$_7$=1 states alone have a poor linear relation, suggesting that these hot cores are not dense enough. This suggests that the excitation of HC$_3$N* requires a dense environment.

\subsection{Differentiation of the HC$_3$N* column density}\label{cdd}

\begin{figure}
\centering
\includegraphics[width=3.2 in,  height=2.6 in]{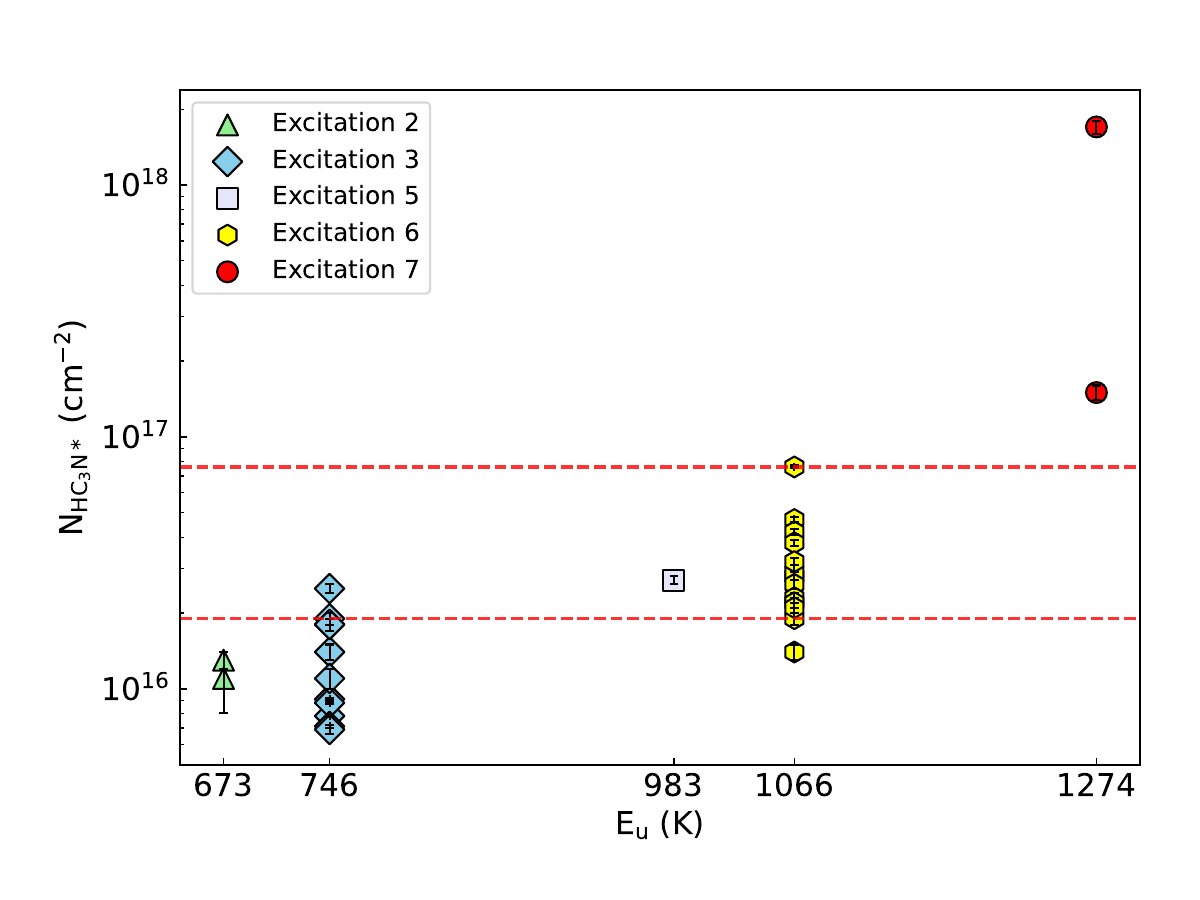}
\caption{Scatter plot of the distribution of the HC$_3$N* column density as a function of upper level energy. The different markers represent different types of hot cores with different numbers of HC$_3$N* emission lines. The two horizontal dashed red lines represent the two distinguishing values for the column density of 1.9$\times$10$^{16}$ cm$^{-2}$ and 7.6$\times$10$^{16}$ cm$^{-2}$, respectively.
\label{fig:UC}}
\end{figure}

Excitation types 2 and 3, which have similar upper energy levels, display similar HC$_3$N* column densities that range from (6.9$\pm$0.3)$\times$10$^{15}$ to (2.5$\pm$0.1)$\times$10$^{16}$ cm$^{-2}$. Similarly, the HC$_3$N* column density values of excitation types 5 and 6, also with comparable upper energy levels, range from (1.4$\pm$0.1)$\times$10$^{16}$ to (7.6$\pm$0.1)$\times$10$^{16}$ cm$^{-2}$. In particular, the two hot cores classified as excitation type 7 exhibit the highest HC$_3$N* column densities. They exceed 10$^{17}$ cm$^{-2}$. Therefore, significant differences in HC$_3$N* column density are observed between these various types of hot cores. Figure~\ref{fig:UC} illustrates the column densities for all types of hot cores with different numbers of HC$_3$N* states. The HC$_3$N* column density clearly increases in general with increasing upper energy level, and hot cores with similar upper energy levels exhibit similar HC$_3$N* column densities.

Based on these excitation types, we identify two distinguishing values of HC$_3$N* column density: 1.9$\times$10$^{16}$ cm$^{-2}$ and 7.6$\times$10$^{16}$ cm$^{-2}$. The column densities of hot cores in excitation types 2 and 3 are for 92.3\% below the first distinguishing value of 1.9$\times$10$^{16}$ cm$^{-2}$. In contrast, the column densities of 92.9\% of the hot cores in excitation types 5 and 6 lie between the two distinguishing values and range from 1.9$\times$10$^{16}$ cm$^{-2}$ to 7.6$\times$10$^{16}$ cm$^{-2}$. In all hot cores in excitation type 7, $\rm N_{H_2}$ exceeds the second distinguishing value of 7.6$\times$10$^{16}$ cm$^{-2}$. Our observations indicate that the number of vibrationally excited states increases with the overall column density of HC$_3$N*, and higher-energy lines are more easily detected when the column density is higher.

\subsection{Excitation factors}

\begin{figure}
\centering
\includegraphics[width=3.2 in,  height=2.6 in]{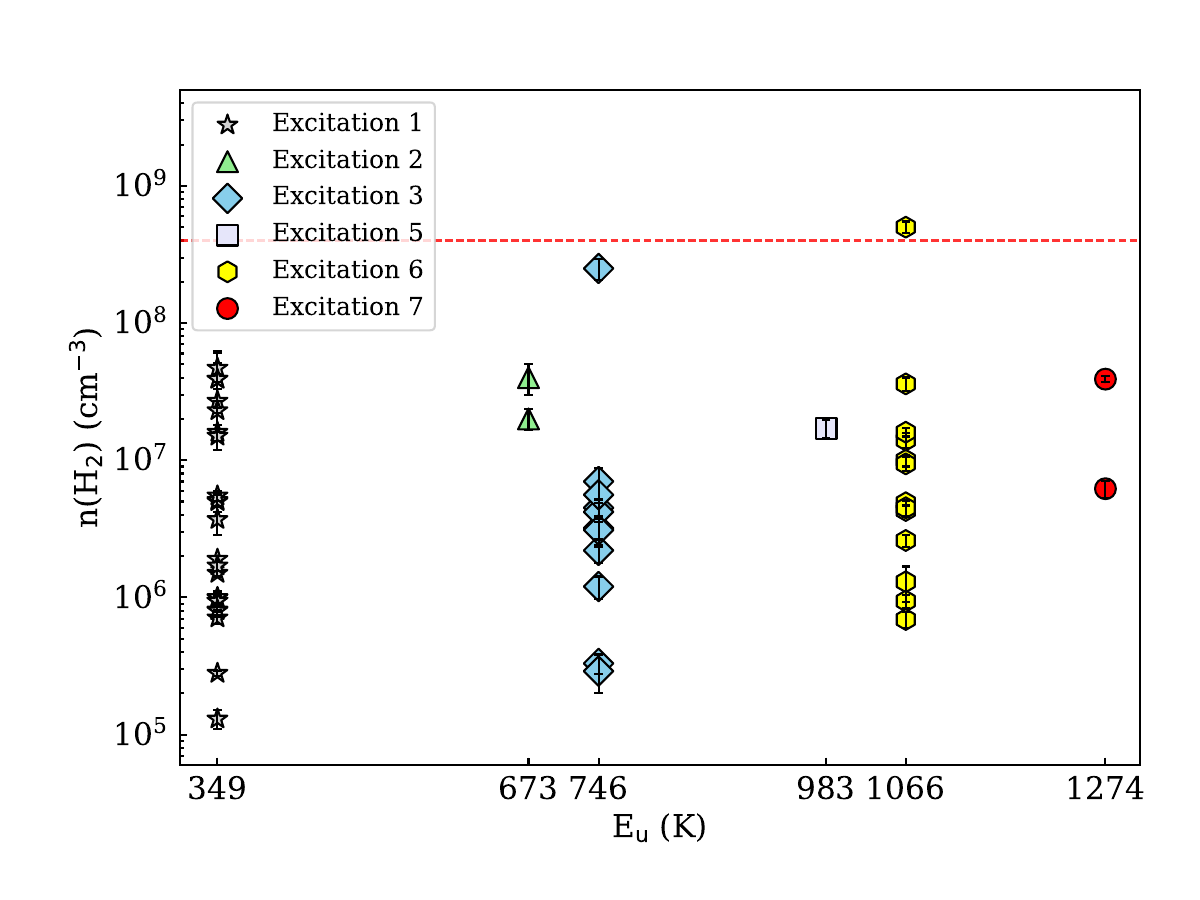}
\caption{Scatter plot of the distribution of H$_2$ number density as a function of upper level energy. The different markers represent different types of hot cores with different numbers of HC$_3$N* emission lines. The horizontal dashed red line represents the critical density of 4$\times$10$^8$ cm$^{-3}$ for HC$_3$N v$_7$=1 excitation. \label{fig:UH}}
\end{figure}

The excitation of HC$_3$N* states can be produced by two different mechanisms: pumping by the absorption of mid-IR photons, and/or collisions with H$_2$. The strong thermal emission by hot dust, which provides high-energy mid-IR photons, can excite HC$_3$N* from a low to a high energy level. The excitation of HC$_3$N* states may also be impacted by the H$_2$ number density n(H$_2$). When n(H$_2$) exceeds n$\rm_{crit}$ for the collision excitation of vibration levels, the excitation of a vibrationally excited state is dominated by collisions with H$_2$. Column 11 in Tables~\ref{tab:param1} to~\ref{tab:param4} lists n(H$_2$) for all sources. Figure~\ref{fig:UH} illustrates the distribution of n(H$_2$) as a function of the upper level energy. All hot cores in our sample except for I18507+0110, for which n(H$_2$) is lower than n$\rm_{crit}$, indicate that excitation of HC$_3$N* is dominated by mid-IR pumping and collisional excitation is relatively ineffective. This agrees with previous observations \citep{1985ApJ...299..405G, 1992A&A...261L...5S}.


\section{Conclusions} \label{con}

We have performed a systematic survey of vibrationally excited HC$_3$N lines (HC$_3$N*) in hot cores using the data of ALMA band 3 survey obtained by the ATOMS project. The main results are summarized below.

(1) Emission from seven different HC$_3$N* states was detected in 52 out of the 60 hot cores. The detection rates for different HC$_3$N* states decrease with increasing upper level energy. When a high-energy level line is detected in a hot core, a low-energy level line is invariably present.

(2) By analyzing the spatial distribution of HC$_3$N v$_7$=1 and UC H{\sc ii} regions, we found that the spatial distribution of HC$_3$N* is influenced by the presence of UC H{\sc ii} regions.

(3) The derived rotation temperature for HC$_3$N* ranges from (160$\pm$1) to (335$\pm$2) K, with a mean value of 235 K, and the H$_2$ number density ranges from (1.3$\pm$0.2)$\times$10$^{5}$ to (5.0$\pm$0.5)$\times$10$^{8}$ cm$^{-3}$, with a mean value of 3.2$\times$10$^{7}$ cm$^{-3}$. The high rotation temperature and substantial H$_2$ number density show that HC$_3$N* is excited from the inner region of hot cores.

(4) The column density of HC$_3$N* ranges from (6.9$\pm$0.3)$\times$10$^{15}$ to (1.7$\pm$0.1)$\times$10$^{18}$ cm$^{-2}$, with a mean value of 2.7$\times$10$^{16}$ cm$^{-2}$. Two distinguishing values of the HC$_3$N* column density for hot cores with different numbers of HC$_3$N* states were obtained. Hot cores with fewer than four HC$_3$N* states have a column density lower than 1.9$\times$10$^{16}$ cm$^{-2}$, while those with more than six HC$_3$N* states have a column density greater than 7.6$\times$10$^{16}$ cm$^{-2}$. The column density of the remaining hot cores falls between these two values.

(5) In hot cores, the HC$_3$N* states are pumped by the absorption of mid-IR photon and not by collisional excitation.

\section*{Data availability}
The derived data underlying this article are available in the article and in its online supplementary material on \href{https://zenodo.org/doi/10.5281/zenodo.14557138}{Zenodo}.

\begin{acknowledgements}
This work has been supported by the National Key R\&D Program of China (No.\ 2022YFA1603100), the National Natural Science Foundation of China (NSFC) through grant Nos. 12033005, 12073061, and 12122307, and the Tianchi Talent Program of Xinjiang Uygur Autonomous Region.
This research was carried out in part at the Jet Propulsion Laboratory, California Institute of Technology, under a contract with the National Aeronautics and Space Administration (80NM0018D0004).
G.G. gratefully acknowledges support by the ANID BASAL project FB210003.
M.Y.T. acknowledges the support by the NSFC through grant No. 12203011, and the Yunnan Provincial Department of Science and Technology through grant No. 202101BA070001-261.
PS was partially supported by a Grant-in-Aid for Scientific Research (KAKENHI Number JP22H01271 and JP23H01221) of JSPS. PS was supported by Yoshinori Ohsumi Fund (Yoshinori Ohsumi Award for Fundamental Research).
SRD acknowledges support from the Fondecyt Postdoctoral fellowship (project code 3220162) and ANID BASAL project FB210003.
LB gratefully acknowledges support by the ANID BASAL project FB210003.
CWL was supported by the Basic Science Research Program through the NRF funded by the Ministry of Education, Science and Technology (NRF- 2019R1A2C1010851) and by the Korea Astronomy and Space Science Institute grant funded by the Korea government (MSIT; project No. 2024-1-841-00).
H.-L. Liu was supported by Yunnan Fundamental Research Project (grant Nos 202301AT070118, and 202401AS070121), and by Xingdian Talent Support Plan–Youth Project.
X.H.L. acknowledges support from the Natural Science Foundation of Xinjiang Uygur Autonomous Region (No. 2024D01E37) and the National Science Foundation of China (12473025).
This work is sponsored in part by the Chinese Academy of Sciences (CAS), through a grant to the CAS South America Center for Astronomy (CASSACA) in Santiago, Chile. This paper makes use of the following ALMA data: ADS/JAO.ALMA\#2019.1.00685.S. ALMA is a partnership of ESO (representing its member states), NSF (USA), and NINS (Japan), together with NRC (Canada), MOST and ASIAA (Taiwan), and KASI (Republic of Korea), in cooperation with the Republic of Chile. The Joint ALMA Observatory is operated by ESO, AUI/NRAO, and NAOJ.
\end{acknowledgements}

\bibliography{hc3n.bib}
\bibliographystyle{aa}

\appendix
\onecolumn

\section{Parameters of vibrationally excited HC$_3$N lines in SPW 8}
\begin{table*}[htp!]
\centering
\caption{Parameters of vibrationally excited HC$_3$N lines.}
\label{tab:HC3N}
\resizebox{15cm}{!}{
\begin{tabular}{cccccccc} 
\hline\hline
State & Frequency & Uncertainty & $\rm J_{up}-J_{low}$ & Parity & $\rm S_{\rm ij}$$\rm \mu^2$ & $\rm Log_{\rm 10}(A_{\rm ij})$ & E$_{\rm u}$ \\
& (MHz) & (MHz) & & & (D$^{\rm 2}$) & ($\rm s^{-1}$) & (K) \\
\hline
v$_4$=1 & 99938.6314 & 0.0040 & $\rm J = 11-10$ & -- & 147.65946 & $-$4.12733 & 1274.35207 \\
v$_5$=1 & 100085.1370 & 0.0055 & $\rm J = 11-10$  & $1e$ & 150.47652 & $-$4.11721 & 983.02211 \\
v$_5$=1 & 100141.3008 & 0.0036 & $\rm J = 11-10$  & $1f$ & 150.47652 & $-$4.11647 & 983.03920 \\
v$_6$=1 & 100240.5843 & 0.0016 & $\rm J = 11-10$  & $1e$ & 151.47591 & $-$4.11232 & 746.53960 \\
v$_6$=1 & 100319.3816 & 0.0016 & $\rm J = 11-10$  & $1f$ & 151.45767 & $-$4.11135 & 746.56237 \\
v$_7$=1 & 100322.4109 & 0.0017 & $\rm J = 11-10$  & $1e$ & 151.18087 & $-$4.11210 & 349.73387 \\
v$_7$=1 & 100466.1745 & 0.0017 & $\rm J = 11-10$  & $1f$ & 151.14994 & $-$4.11033 & 349.77530 \\
v$_6$=v$_7$=1 & 100593.4944 & 0.0030 & $\rm J = 11-10$  & $0^{+}$ & 151.98949 & $-$4.10627 & 1065.29153 \\
v$_6$=v$_7$=1 & 100596.2397 & 0.0032 & $\rm J = 11-10$  & $0^{-}$ & 151.95748 & $-$4.10633 & 1066.4191 \\
v$_6$=v$_7$=1 & 100604.0227 & 0.0024 & $\rm J = 11-10$  & $2^{-}$ & 146.97639 & $-$4.12070 & 1066.62147 \\
v$_6$=v$_7$=1 & 100606.2047 & 0.0026 & $\rm J = 11-10$  & $2^{+}$ & 146.97012 & $-$4.12069 & 1066.62172 \\
v$_7$=2 & 100708.7840 & 0.0020 & $\rm J = 11-10$  & $0$ & 151.73124 & $-$4.10552 & 670.68918 \\
v$_7$=2 & 100711.0640 & 0.0015 & $\rm J = 11-10$  & $2e$ & 146.72153 & $-$4.12007 & 673.96249 \\
v$_7$=2 & 100714.3951 & 0.0016 & $\rm J = 11-10$  & $2f$ & 146.71208 & $-$4.12005 & 673.96308 \\
v$_7$=3 & 100880.5889 & 0.0049 & $\rm J = 11-10$  & $1e$ & 149.75614 & $-$4.10899 & 983.88413 \\
v$_7$=3 & 101027.8491 & 0.0031 & $\rm J = 11-10$  & $3$ & 139.77879 & $-$4.13703 & 990.39731 \\
v$_7$=3 & 101027.8709 & 0.0031 & $\rm J = 11-10$  & $3$ & 139.77873 & $-$4.13703 & 990.39731 \\
v$_7$=3 & 101169.9207 & 0.0040 & $\rm J = 11-10$  & $1f$ & 149.72959 & $-$4.10533 & 983.96751 \\
\hline
\end{tabular}}
\begin{flushleft}
{ NOTE.} The rest frequencies for all states are listed with uncertainties, and all transitions have the same $\rm J=11-10$.
\end{flushleft}
\end{table*}

\section{Physical parameters derived from vibrationally exited HC$_3$N lines}

Table~\ref{tab:catalog} lists the detected HC$_3$N* states in hot cores, along with their corresponding rotation temperatures and column densities.

\section{The basic physical parameters of hot cores}

Tables~\ref{tab:param1} to \ref{tab:param4} present the basic physical parameters of the hot cores. These tables categorize and list the parameters of hot cores based on whether they are associated with UC H{\sc ii} regions and show HC$_3$N* emission.

\section{Spectra of vibrationally excited HC$_3$N lines in SPW 8}

In the survey observations, we detected a total of 7 types of HC$_3$N* lines in 52 hot cores, comprising 18 transitions in SPW 8, with none detected in SPW 7, all corresponding to the same J=11$-$10 transition. Of the 52 hot cores, 29 show more than one HC$_3$N* state, and we performed XCLASS fitting on the spectral lines. In Fig. D.1 (on \href{https://zenodo.org/doi/10.5281/zenodo.14557138}{Zenodo}), we labeled all detected HC$_3$N* transitions and one HC$_3$N v=0 ground state transition.

\section{Rotational temperature diagram}

Figure E.1 (on \href{https://zenodo.org/doi/10.5281/zenodo.14557138}{Zenodo}) presents the RTD analysis for 29 hot cores. Since XCLASS considers the effect of optical depth on line flux, while our RTD calculations do not apply this correction, the rotation temperature derived from RTD is unreliable for optically thick sources. Of these 29 hot cores, only IRAS 18056$-$1952 is optically thick. For the remaining 28 hot cores, the rotation temperatures obtained from XCLASS and RTD are consistent when considering the errors.

\begin{table*}
\renewcommand{\thetable}{B.1}
\centering
\caption{Physical parameters derived from vibrationally exited HC$_3$N lines.}
\label{tab:catalog}
\begin{adjustbox}{width=0.88\textwidth}
\begin{tabular}{lcccccccccc} 
   \hline\hline
Source & v$_7$=1 & v$_7$=2 & v$_6$=1 & v$_5$=1 & v$_7$=3 & v$_6$=v$_7$=1 & v$_4$=1 & Excitation & T$_{\rm rot}$ (K) & N (cm$^{-2}$) \\  
\hline
I08303--4303 & $\times$ & $\times$ &$\times$&$\times$ & $\times$ & $\times$ &$\times$ & 0 & $\cdots$& $\cdots$ \\
I08470--4243 & $\checkmark$ & $\times$ &$\times$&$\times$ & $\times$ & $\times$ &$\times$ & 1 & $\cdots$& $\cdots$ \\
I09018--4816 & $\times$ & $\times$ &$\times$&$\times$ & $\times$ & $\times$ &$\times$ & 0 & $\cdots$& $\cdots$ \\
I11298--6155 & $\times$ & $\times$ &$\times$&$\times$ & $\times$ & $\times$ &$\times$ & 0 & $\cdots$& $\cdots$ \\
I12326--6245 &$\checkmark$ &$\checkmark$  &$\times$& $\times$ &$\times$ &$\times$ &$\times$& 2 & 189$\pm$16 & (1.3$\pm$0.1)$\times$10$^{16}$ \\
I13079--6218 & $\checkmark$ & $\checkmark$ &$\checkmark$& $\checkmark$ & $\checkmark$ & $\checkmark$ &$\times$ & 6 & 253$\pm$14 &  (2.8$\pm$0.1)$\times$10$^{16}$ \\
I13134--6242 & $\checkmark$ &$\checkmark$ &$\checkmark$& $\checkmark$ & $\checkmark$ & $\checkmark$ &$\times$ & 6 & 335$\pm$22 &  (1.9$\pm$0.1)$\times$10$^{16}$ \\
I13140--6226 &$\checkmark$ & $\times$ &$\times$&$\times$ & $\times$ & $\times$ &$\times$ & 1 & $\cdots$& $\cdots$ \\
I13471--6120 &$\checkmark$ & $\times$ &$\times$&$\times$ & $\times$ & $\times$ &$\times$ & 1 & $\cdots$& $\cdots$ \\
I13484--6100 &$\checkmark$ &$\times$ &$\times$&$\times$ & $\times$ & $\times$ &$\times$ & 1 & $\cdots$& $\cdots$ \\
I14498--5856 &$\checkmark$ & $\times$ &$\times$&$\times$ & $\times$ & $\times$ &$\times$ & 1 & $\cdots$& $\cdots$ \\
I15254--5621 &$\checkmark$ &$\times$ &$\times$&$\times$ & $\times$ &$\times$ &$\times$ & 1 & $\cdots$& $\cdots$ \\
I15437--5343 &$\checkmark$ &$\times$ &$\times$&$\times$ & $\times$ & $\times$ &$\times$ & 1 & $\cdots$& $\cdots$ \\
I15520--5234  &$\checkmark$ & $\times$ &$\times$&$\times$ & $\times$ & $\times$ &$\times$ & 1 & $\cdots$& $\cdots$ \\
I16060--5146  &$\checkmark$ &$\checkmark$  & $\checkmark$ & $\times$ & $\times$ & $\times$ & $\times$ & 3 & 307$\pm$33 &(9.1$\pm$0.1)$\times$10$^{15}$  \\
I16065--5158  &$\checkmark$ &$\checkmark$ &$\checkmark$& $\checkmark$ & $\checkmark$ & $\checkmark$ & $\times$ & 6 & 248$\pm$21 & (4.7$\pm$0.1)$\times$10$^{16}$ \\
I16071--5142  &$\checkmark$ & $\checkmark$  & $\checkmark$ & $\times$ & $\times$ & $\times$ & $\times$  & 3 & 161$\pm$18 & (1.4$\pm$0.1)$\times$10$^{16}$ \\
I16076--5134  &$\checkmark$ & $\checkmark$ & $\checkmark$ & $\times$ & $\times$ & $\times$ & $\times$ & 3  & 160$\pm$12 & (7.8$\pm$0.1)$\times$10$^{15}$ \\
I16164--5046  &$\checkmark$ & $\checkmark$ & $\checkmark$ & $\times$ & $\times$ & $\times$ & $\times$   & 3 & 262$\pm$34 & (8.8$\pm$0.1)$\times$10$^{15}$ \\
I16172--5028 &$\checkmark$ & $\times$ &$\times$&$\times$ & $\times$ & $\times$ &$\times$ & 1 & $\cdots$& $\cdots$ \\
I16272--4837c1 &$\checkmark$ &$\checkmark$ & $\checkmark$ & $\times$ & $\times$ & $\times$ & $\times$   & 3 & 169$\pm$32 & (1.9$\pm$0.1)$\times$10$^{16}$ \\
I16272--4837c2 &$\checkmark$ & $\times$ & $\times$&$\times$ & $\times$ & $\times$ &$\times$ & 1 & $\cdots$& $\cdots$ \\
I16272--4837c3 &$\checkmark$ &$\times$ & $\times$&$\times$ & $\times$ & $\times$ &$\times$ & 1 & $\cdots$& $\cdots$ \\
I16318--4724 &$\checkmark$ & $\checkmark$ &$\checkmark$& $\checkmark$ & $\checkmark$ & $\checkmark$ &$\times$ & 6  &229$\pm$12 & (2.9$\pm$0.2)$\times$10$^{16}$ \\
I16344--4658  &$\checkmark$ & $\times$ &$\times$&$\times$ & $\times$ & $\times$ &$\times$ & 1 & $\cdots$& $\cdots$ \\
I16348--4654  &$\checkmark$ & $\checkmark$ &$\checkmark$& $\checkmark$ & $\checkmark$ & $\checkmark$ &$\checkmark$  & 7 & 313$\pm$12 & (1.5$\pm$0.1)$\times$10$^{17}$ \\
I16351--4722  & $\times$ & $\times$ &$\times$&$\times$ & $\times$ & $\times$ &$\times$ & 0 & $\cdots$& $\cdots$ \\
I16458--4512  & $\times$ & $\times$ &$\times$&$\times$ & $\times$ & $\times$ &$\times$ & 0 & $\cdots$& $\cdots$ \\
I16484--4603 &$\checkmark$ & $\times$ & $\times$ & $\times$ & $\times$ & $\times$ &$\times$ & 1 & $\cdots$& $\cdots$ \\
I16547--4247  &$\checkmark$ & $\checkmark$ & $\checkmark$& $\checkmark$ & $\checkmark$ & $\checkmark$ &$\times$ & 6 & 277$\pm$32 & (2.6$\pm$0.1)$\times$10$^{16}$ \\
I17008--4040 &$\checkmark$ & $\checkmark$ &$\checkmark$& $\checkmark$ & $\checkmark$ & $\checkmark$ & $\times$ & 6 & 224$\pm$14 & (2.3$\pm$0.1)$\times$10$^{16}$ \\
I17016--4124c1  & $\checkmark$ & $\checkmark$ & $\checkmark$ & $\times$ & $\times$   & $\times$  & $\times$ & 3 & 190$\pm$20 & (2.5$\pm$0.1)$\times$10$^{16}$ \\
I17016--4124c2 & $\times$ & $\times$ &$\times$&$\times$ & $\times$ & $\times$ &$\times$ & 0 & $\cdots$& $\cdots$ \\
I17158--3901c1 & $\checkmark$ & $\times$ &$\times$&$\times$ & $\times$ & $\times$ &$\times$ & 1 & $\cdots$& $\cdots$ \\
I17158--3901c2  & $\checkmark$ & $\times$ &$\times$&$\times$ & $\times$ & $\times$ &$\times$ & 1 & $\cdots$& $\cdots$ \\
I17175--3544  & $\checkmark$ &$\checkmark$ &$\checkmark$& $\checkmark$ & $\checkmark$ & $\checkmark$ & $\times$ & 6 & 321$\pm$8 & (4.2$\pm$0.1)$\times$10$^{16}$ \\
I17220--3609  & $\checkmark$ &$\checkmark$ & $\checkmark$ & $\times$ & $\times$ & $\times$ & $\times$  & 3 & 165$\pm$15 & (1.8$\pm$0.1)$\times$10$^{16}$ \\
I17233--3606  & $\checkmark$ & $\checkmark$ &$\checkmark$& $\checkmark$ & $\checkmark$ & $\times$ & $\times$ & 5 & 185$\pm$9 & (2.7$\pm$0.1)$\times$10$^{16}$ \\
I17441--2822  &$\checkmark$ & $\checkmark$ &$\checkmark$& $\checkmark$ & $\checkmark$ & $\checkmark$ &$\times$ & 6 & 232$\pm$28 & (3.8$\pm$0.1)$\times$10$^{16}$ \\
I18032--2032c1 & $\times$ & $\times$ &$\times$&$\times$ & $\times$ & $\times$ &$\times$ & 0 & $\cdots$& $\cdots$ \\
I18032--2032c2 &$\checkmark$ & $\checkmark$ &$\checkmark$& $\checkmark$ & $\checkmark$ & $\checkmark$ &$\times$ & 6 & 210$\pm$7 & (1.4$\pm$0.1)$\times$10$^{16}$ \\
I18032--2032c3 &$\checkmark$ & $\times$ &$\times$&$\times$ & $\times$ & $\times$ &$\times$ & 1 & $\cdots$& $\cdots$ \\
I18032--2032c4 &$\checkmark$ & $\times$ &$\times$&$\times$ & $\times$ & $\times$ &$\times$ & 1 & $\cdots$& $\cdots$ \\
I18056--1952  &$\checkmark$ & $\checkmark$ &$\checkmark$& $\checkmark$ & $\checkmark$ & $\checkmark$ &$\checkmark$ & 7 & 270$\pm$5 & (1.7$\pm$0.1)$\times$10$^{18}$ \\
I18089--1732 &$\checkmark$ &$\checkmark$ &$\checkmark$& $\checkmark$ & $\checkmark$ & $\checkmark$ &$\times$ & 6 & 276$\pm$37 & (3.2$\pm$0.1)$\times$10$^{16}$ \\
I18117--1753  &$\checkmark$ & $\checkmark$  & $\checkmark$ & $\times$ & $\times$  & $\times$  &$\times$ & 3 & 165$\pm$8 & (1.8$\pm$0.1)$\times$10$^{16}$ \\
I18159--1648c1 &$\checkmark$ &$\times$ &$\times$&$\times$ & $\times$ & $\times$ &$\times$ & 1 & $\cdots$& $\cdots$ \\
I18159--1648c2 &$\checkmark$ &$\times$ &$\times$&$\times$ & $\times$ & $\times$ &$\times$ & 1 & $\cdots$& $\cdots$ \\
I18182--1433  &$\checkmark$ &$\times$ &$\times$&$\times$ & $\times$ & $\times$ &$\times$ & 1 & $\cdots$& $\cdots$ \\
I18236--1205  &$\checkmark$ &$\times$ &$\times$&$\times$ & $\times$ & $\times$ &$\times$ & 1 & $\cdots$& $\cdots$ \\ 
I18290--0924  & $\times$ & $\times$ &$\times$&$\times$ & $\times$ & $\times$ &$\times$ & 0 & $\cdots$& $\cdots$ \\
I18316--0602 & $\checkmark$ & $\checkmark$ & $\checkmark$ & $\times$ & $\times$ & $\times$ & $\times$ & 3 & 244$\pm$22 & (7.1$\pm$0.1)$\times$10$^{15}$ \\
I18411--0338  &$\checkmark$ &$\checkmark$ &$\checkmark$& $\checkmark$ & $\checkmark$ & $\checkmark$ &$\times$ & 6 & 273$\pm$7 & (2.2$\pm$0.1)$\times$10$^{16}$ \\ 
I18469--0132  &$\checkmark$ &$\checkmark$ &$\times$&$\times$  & $\times$  & $\times$  &$\times$ & 2 & 239$\pm$67 & (1.1$\pm$0.3)$\times$10$^{16}$ \\  
I18507+0110   &$\checkmark$ &$\checkmark$ &$\checkmark$& $\checkmark$ &$\checkmark$ & $\checkmark$ &$\times$ & 6 & 245$\pm$19 &(7.6$\pm$0.1)$\times$10$^{16}$ \\
I18507+0121 &$\checkmark$ &$\checkmark$ &$\checkmark$& $\checkmark$ &$\checkmark$ & $\checkmark$ &$\times$ & 6 & 218$\pm$20 &(2.1$\pm$0.1)$\times$10$^{16}$ \\ 
I18517+0437  &$\checkmark$ &$\times$ &$\times$&$\times$ & $\times$ & $\times$ &$\times$ & 1 &  $\cdots$& $\cdots$ \\ 
I19078+0901c1  &$\checkmark$ &$\checkmark$ & $\checkmark$ & $\times$ & $\times$  &$\times$ &$\times$ & 3 & 235$\pm$48 & (6.9$\pm$0.3)$\times$10$^{15}$ \\   
I19078+0901c2 &$\checkmark$ &$\checkmark$ & $\checkmark$ & $\times$ & $\times$  &$\times$ &$\times$ & 3 & 217$\pm$14 & (1.1$\pm$0.1)$\times$10$^{16}$ \\   
I19095+0930  & $\checkmark$ &$\times$ &$\times$&$\times$ & $\times$ & $\times$ &$\times$ & 1 & $\cdots$& $\cdots$ \\ 
\hline
\end{tabular}
\end{adjustbox}
\begin{flushleft}
{Notes:} Column 1 lists the source names. From column 2 to column 8, the $\checkmark$ marks the existence and the $\times$ marks the inexistence of a HC$_3$N* state. In column 9, the number of Excitation means the number of excited HC$_3$N* states for each hot core. Columns 10 and 11 list the fitted rotation temperature and column density for HC$_3$N* states.
\end{flushleft}
\end{table*}

\begin{sidewaystable*}
\renewcommand{\thetable}{C.1}
\centering
\caption{Basic physical parameters of hot cores without UC H{\sc ii} regions which show HC$_3$N* emission.}
\label{tab:param1}
\begin{adjustbox}{width=\textwidth}
\begin{tabular}{lccccccccccc}
\hline\hline
Source & Distance & R$\rm_{GC}$ & Resolution & R$\rm_{core}$ & T$ \rm _{d}\:^{a}$ & S$ \rm _{\nu} ^{int}$ & S$ \rm _{\nu} ^{peak}$ & M$ \rm _{core}$ & $ \rm N_{H_2}$ & n(H$_2$) & $ \rm f_{HC_3N*}$\\
 & (kpc) & (kpc) & (au) & (au) & (K) & (mJy) & (mJy beam$^{-1}$) & (M$_{\sun}$) & (cm$^{-2}$) & (cm$^{-3}$) & \\
\hline
I08470--4243 & 2.10 & 8.8 & 2520 & 3780 & 200$\pm$30 & 37.8$\pm$2.3 & 18.8$\pm$0.7 & 9.0$\pm$1 & (3.8$\pm$0.6)$\times$10$^{23}$ & (5.0$\pm$0.8)$\times$10$^{6}$ & $\cdots$ \\
I13079--6218 & 3.80 & 6.9 & 4560 & 9120 & 110$\pm$9 & 109.5$\pm$2.9 & 62.5$\pm$0.9 & 106.8$\pm$9 & (7.8$\pm$0.7)$\times$10$^{23}$ & (4.3$\pm$0.4)$\times$10$^{6}$ & (3.6$\pm$0.3)$\times$10$^{-8}$ \\
I13134--6242 & 3.80 & 6.9 & 6460 & 8360 & 160$\pm$5 & 143.1$\pm$2.3 & 86.1$\pm$0.9 & 95.3$\pm$4 & (8.2$\pm$0.3)$\times$10$^{23}$ & (4.9$\pm$0.2)$\times$10$^{6}$ & (2.3$\pm$0.2)$\times$10$^{-8}$ \\
I13140--6226 & 3.80 & 6.9 & 6460 & 20900 & 120$\pm$17 & 44$\pm$2.4 & 12.7$\pm$0.6 & 39.3$\pm$6 & (5.4$\pm$0.8)$\times$10$^{22}$ & (1.3$\pm$0.2)$\times$10$^{5}$ & $\cdots$ \\
I13484--6100 & 5.40 & 6.4 & 9720 & 13500 & 131$\pm$18 & 51.2$\pm$1.7 & 23.7$\pm$0.6 & 76.3$\pm$11 & (2.5$\pm$0.4)$\times$10$^{23}$ & (9.4$\pm$1.3)$\times$10$^{5}$ & $\cdots$ \\
I14498--5856 & 3.16 & 6.4 & 5688 & 8848 & 106$\pm$3 & 67.3$\pm$3.1 & 27.4$\pm$0.9 & 42.6$\pm$2 & (3.3$\pm$0.2)$\times$10$^{23}$ & (1.9$\pm$0.1)$\times$10$^{6}$ & $\cdots$ \\
I15437--5343 & 4.98 & 5.0 & 8466 & 11454 & 106$\pm$11 & 29.6$\pm$0.5 & 15.1$\pm$0.2 & 35.2$\pm$4 & (1.6$\pm$0.2)$\times$10$^{23}$ & (7.1$\pm$0.8)$\times$10$^{5}$ & $\cdots$ \\
I16272--4837c1 & 2.92 & 5.8 & 4964 & 6424 & 230$\pm$37 & 127$\pm$6.4 & 63.9$\pm$2.2 & 27.7$\pm$5 & (4.1$\pm$0.7)$\times$10$^{23}$ & (3.2$\pm$0.5)$\times$10$^{6}$ & (4.7$\pm$0.8)$\times$10$^{-8}$ \\
I16272--4837c2 & 2.92 & 5.8 & 4964 & 5840 & 150$\pm$9 & 19.3$\pm$0.9 & 9.5$\pm$0.3 & 6.5$\pm$1 & (1.2$\pm$0.1)$\times$10$^{23}$ & (9.9$\pm$0.8)$\times$10$^{5}$ & $\cdots$ \\
I16272--4837c3 & 2.92 & 5.8 & 4964 & 6132 & 123$\pm$12 & 19.1$\pm$0.7 & 9.5$\pm$0.3 & 7.9$\pm$1 & (1.3$\pm$0.1)$\times$10$^{23}$ & (1.0$\pm$0.1)$\times$10$^{6}$ & $\cdots$ \\
I16318--4724 & 7.68 & 3.3 & 13056 & 16896 & 148$\pm$15 & 104$\pm$3.2 & 49.4$\pm$1.1 & 148.9$\pm$16 & (3.2$\pm$0.3)$\times$10$^{23}$ & (9.4$\pm$1.0)$\times$10$^{5}$ & (9.2$\pm$1.2)$\times$10$^{-8}$ \\
I16344--4658 & 12.09 & 5.4 & 20553 & 20553 & 160$\pm$15 & 54.2$\pm$1.4 & 31.5$\pm$0.7 & 270.5$\pm$27 & (3.9$\pm$0.4)$\times$10$^{23}$ & (9.4$\pm$0.9)$\times$10$^{5}$ & $\cdots$ \\
I16484--4603 & 2.10 & 6.4 & 3570 & 2730 & 127$\pm$5 & 65.0$\pm$1.0 & 45.0$\pm$0.4 & 15.1$\pm$1 & (1.2$\pm$0.1)$\times$10$^{24}$ & (2.3$\pm$0.1)$\times$10$^{7}$ & $\cdots$ \\
I16547--4247 & 2.74 & 5.8 & 5206 & 6850 & 150$\pm$20 & 160.8$\pm$6.1 & 74.8$\pm$2.1 & 47.7$\pm$7 & (6.1$\pm$0.9)$\times$10$^{23}$ & (4.5$\pm$0.6)$\times$10$^{6}$ & (4.2$\pm$0.6)$\times$10$^{-8}$ \\
I17008--4040 & 2.38 & 6.1 & 4522 & 4046 & 150$\pm$14 & 94.2$\pm$3.4 & 58.7$\pm$1.2 & 22.4$\pm$2 & (8.3$\pm$0.8)$\times$10$^{23}$ & (1.0$\pm$0.1)$\times$10$^{7}$ & (2.8$\pm$0.3)$\times$10$^{-8}$ \\
I17158--3901c1 & 3.38 & 5.1 & 5746 & 9802 & 151$\pm$12 & 64.6$\pm$2.3 & 54.6$\pm$0.7 & 25.2$\pm$2 & (1.6$\pm$0.1)$\times$10$^{23}$ & (8.1$\pm$0.7)$\times$10$^{5}$ & $\cdots$ \\
I17158--3901c2 & 3.38 & 5.1 & 5746 & 9464 & 152$\pm$3 & 20.3$\pm$0.7 & 10.2$\pm$0.2 & 7.9$\pm$1 & (5.3$\pm$0.2)$\times$10$^{22}$ & (2.8$\pm$0.1)$\times$10$^{5}$ & $\cdots$ \\
I17233--3606 & 1.34 & 7.0 & 2278 & 5226 & 100$\pm$14 & 596$\pm$22 & 130.9$\pm$3.7 & 81.3$\pm$12 & (1.8$\pm$0.3)$\times$10$^{24}$ &(1.7$\pm$0.3)$\times$10$^{7}$ & (1.5$\pm$0.2)$\times$10$^{-8}$ \\
I18032--2032c2 & 5.15 & 3.4 & 7725 & 12360 & 130$\pm$12 & 57.1$\pm$5.9 & 17.8$\pm$1.4 & 42.8$\pm$6 & (1.7$\pm$0.2)$\times$10$^{23}$ & (6.9$\pm$0.1)$\times$10$^{5}$ & (8.3$\pm$1.3)$\times$10$^{-8}$ \\
I18032--2032c3 & 5.15 & 3.4 & 7725 & 5665 & 110$\pm$3 & 34.7$\pm$3 & 7.4$\pm$0.6 & 30.8$\pm$3 & (5.8$\pm$0.5)$\times$10$^{23}$ &(5.1$\pm$0.5)$\times$10$^{6}$ & $\cdots$ \\
I18056--1952 & 8.55 & 1.6 & 12825 & 9405 & 133$\pm$5 & 767$\pm$8.4 & 600.3$\pm$4.3 & 1079.2$\pm$53 & (7.4$\pm$0.4)$\times$10$^{24}$ &(3.9$\pm$0.2)$\times$10$^{7}$ & (2.3$\pm$0.2)$\times$10$^{-7}$ \\
I18089--1732 & 2.50 & 5.9 & 4250 & 4250 & 128$\pm$14 & 80.1$\pm$2.7 & 45.8$\pm$1.1 & 23.7$\pm$3 & (7.9$\pm$0.9)$\times$10$^{23}$ & (9.4$\pm$1.1)$\times$10$^{6}$ & (4.0$\pm$0.5)$\times$10$^{-8}$ \\
I18117--1753 & 2.57 & 5.9 & 4369 & 4112 & 160$\pm$25 & 38.4$\pm$1.6 & 22.5$\pm$0.5 & 9.6$\pm$2 & (3.4$\pm$0.6)$\times$10$^{23}$ &(4.2$\pm$0.7)$\times$10$^{6}$ & (5.3$\pm$0.9)$\times$10$^{-8}$ \\
I18159--1648c1 & 1.48 & 6.9 & 2516 & 2516 & 161$\pm$18 & 84.9$\pm$6.9 & 36.5$\pm$1.4 & 8.5$\pm$1 & (8.1$\pm$1.1)$\times$10$^{23}$ &(1.6$\pm$0.2)$\times$10$^{7}$ & $\cdots$ \\
I18159--1648c2 & 1.48 & 6.9 & 2516 & 2812 & 110$\pm$22 & 73.8$\pm$5.3 & 21.2$\pm$1.2 & 10.9$\pm$2 & (8.4$\pm$1.8)$\times$10$^{23}$ &(1.5$\pm$0.3)$\times$10$^{7}$ & $\cdots$ \\
I18182--1433 & 4.71 & 4.1 & 8007 & 9891 & 122$\pm$4 & 62.5$\pm$1.6 & 29.2$\pm$0.5 & 48.1$\pm$2 & (3.0$\pm$0.1)$\times$10$^{23}$ &(1.5$\pm$0.1)$\times$10$^{6}$ & $\cdots$ \\
I18236--1205 & 2.17 & 6.3 & 3689 & 3038 & 110$\pm$8 & 18$\pm$0.48 & 11.6$\pm$0.2 & 5.1$\pm$1 & (3.3$\pm$0.3)$\times$10$^{23}$ &(5.5$\pm$0.4)$\times$10$^{6}$ & $\cdots$ \\
I18316--0602 & 2.09 & 6.5 & 3135 & 6061 & 108$\pm$204 & 57.7$\pm$3.2 & 24.8$\pm$1 & 16.0$\pm$3 & (2.6$\pm$0.5)$\times$10$^{23}$ & (2.2$\pm$0.4)$\times$10$^{6}$ & (2.7$\pm$0.5)$\times$10$^{-8}$ \\
I18411--0338 & 7.41 & 4.0 & 11115 & 7410 & 160$\pm$15 & 24.6$\pm$1.1 & 17.7$\pm$0.5 & 34.8$\pm$4 & (3.8$\pm$0.4)$\times$10$^{23}$ &(2.6$\pm$0.3)$\times$10$^{6}$ & (5.7$\pm$0.7)$\times$10$^{-8}$ \\ 
I18507+0121 & 1.56 & 7.1 & 2340 & 2652 & 200$\pm$12 & 103.3$\pm$3 & 61.3$\pm$1.2 & 9.6$\pm$1 & (8.3$\pm$0.6)$\times$10$^{23}$ &(1.6$\pm$0.1)$\times$10$^{7}$ & (2.5$\pm$0.2)$\times$10$^{-8}$ \\ 
I18517+0437 & 2.36 & 6.6 & 3540 & 5428 & 140$\pm$23 & 32.3$\pm$1.8 & 15.0$\pm$0.7 & 8.9$\pm$2 & (1.8$\pm$0.3)$\times$10$^{23}$ &(1.7$\pm$0.3)$\times$10$^{6}$ & $\cdots$ \\
I19078+0901c2 & 11.11 & 7.6 & 16665 & 19998 & 115$\pm$6 & 90$\pm$21 & 53.8$\pm$8.2 & 824.9$\pm$197 & (1.2$\pm$0.3)$\times$10$^{24}$ &(3.1$\pm$0.9)$\times$10$^{6}$ & (8.8$\pm$2.3)$\times$10$^{-9}$ \\   
\hline
\end{tabular}
\end{adjustbox}
\begin{flushleft}
{$^{a.}${} The dust temperature is adopted from the temperature of COMs (CH$_3$OCHO, C$_2$H$_5$CN, and CH$_3$OH) calculated by \cite{2022MNRAS.511.3463Q}. The temperature selection for each source is: prioritize CH$_3$OCHO if available, otherwise select C$_2$H$_5$CN, and choose CH$_3$OH if it is the only one present. The same selection criteria apply to Table~\ref{tab:param2}, Table~\ref{tab:param3}, and Table~\ref{tab:param4}.}
\end{flushleft}
\end{sidewaystable*}

\begin{landscape}
\renewcommand{\thetable}{C.2}
\begin{table}
\centering
\caption{Basic physical parameters of hot cores without UC H{\sc ii} regions 
without HC$_3$N* emission.}
\label{tab:param2}
\resizebox{1.34\textwidth}{!}{
\begin{tabular}{lccccccccccc}
\hline\hline
Source & Distance & R$\rm_{GC}$ & Resolution & R$\rm_{core}$ & T$ \rm _{d}$ & S$ \rm _{\nu} ^{int}$ & S$ \rm _{\nu} ^{peak}$ & M$ \rm _{core}$ & $ \rm N_{H_2}$ & n(H$_2$) \\
 & (kpc) & (kpc) & (au) & (au) & (K) & (mJy) & (mJy beam$^{-1}$) & (M$_{\sun}$) & (cm$^{-2}$) & (cm$^{-3}$) \\
\hline
I08303--4303 & 2.30 & 9.0 & 2760 & 5060 & 101$\pm$9 & 29.1$\pm$4.3 & 8.7$\pm$0.3 & 17.3$\pm$3 & (4.1$\pm$0.7)$\times$10$^{23}$ &(4.0$\pm$0.7)$\times$10$^{6}$ \\
I09018--4816 & 2.60 & 8.8 & 3120 & 8840 & 160$\pm$8 & 161$\pm$22 & 13.0$\pm$0.6 & 73.4$\pm$11 & (5.7$\pm$0.8)$\times$10$^{23}$ &(3.2$\pm$0.5)$\times$10$^{6}$ \\
I11298--6155 & 10.00 & 10.1 & 18000 & 31000 & 124$\pm$6 & 42.6$\pm$2.5 & 19.0$\pm$0.8 & 483.4$\pm$37 & (3.0$\pm$0.2)$\times$10$^{23}$ &(4.9$\pm$0.4)$\times$10$^{5}$ \\
I16458--4512 & 3.56 & 5.1 & 6052 & 3916 & 120$\pm$4 & 178.6$\pm$2.8 & 136.1$\pm$1.3 & 97.5$\pm$4 & (3.8$\pm$0.2)$\times$10$^{24}$ &(4.9$\pm$0.2)$\times$10$^{7}$ \\
I18290--0924 & 5.34 & 4.0 & 8010 & 7476 & 120$\pm$4 & 8.0$\pm$0.3 & 6.1$\pm$0.2 & 7.9$\pm$1 & (8.5$\pm$0.5)$\times$10$^{22}$ & (5.7$\pm$0.3)$\times$10$^{5}$ \\
\hline
\end{tabular}}
\end{table}

\begin{table}
\renewcommand{\thetable}{C.3}
\centering
\caption{Basic physical parameters of hot cores associated with UC H{\sc ii} regions
which show HC$_3$N* emission}.
\label{tab:param3}
\resizebox{1.34\textwidth}{!}{
\begin{tabular}{lccccccccccc}
\hline\hline
Source & Distance & R$\rm_{GC}$ & Resolution & R$\rm_{core}$ & T$ \rm _{d}$ & S$ \rm _{\nu} ^{int}$ & S$ \rm _{\nu} ^{peak}$ & M$ \rm _{core}$ & $ \rm N_{H_2}$ & n(H$_2$) &  $ \rm f_{HC_3N*}$ \\
 & (kpc) & (kpc) & (au) & (au) & (K) & (mJy) & (mJy beam$^{-1}$) & (M$_{\sun}$) & (cm$^{-2}$) & (cm$^{-3}$) & \\
\hline
I12326--6245 & 4.61 & 7.2 & 7837 & 8298 & 137$\pm$33 & 620$\pm$53 & 337$\pm$20 & 755.6$\pm$193 & (6.6$\pm$1.7)$\times$10$^{24}$ & (4.0$\pm$1.0)$\times$10$^{7}$ & (2.0$\pm$0.5)$\times$10$^{-9}$ \\
I13471--6120 & 5.46 & 6.4 & 9828 & 5460 & 126$\pm$10 & 121.2$\pm$9.6 & 106.2$\pm$5 & 192.2$\pm$22 & (3.9$\pm$0.4)$\times$10$^{24}$ & (3.6$\pm$0.4)$\times$10$^{7}$ & $\cdots$ \\
I15254--5621 & 4.00 & 5.7 & 7200 & 3200 & 146$\pm$14 & 247$\pm$12 & 201.6$\pm$6.3 & 157.3$\pm$17 & (9.3$\pm$1.0)$\times$10$^{24}$ & (1.5$\pm$0.2)$\times$10$^{8}$ & $\cdots$ \\
I15520--5234 & 2.65 & 6.2 & 4505 & 7685 & 160$\pm$30 & 153$\pm$20 & 106.9$\pm$9.1 & 43.1$\pm$10 & (4.4$\pm$1.0)$\times$10$^{23}$ & (2.9$\pm$0.7)$\times$10$^{6}$ & $\cdots$ \\
I16060--5146 & 5.30 & 4.5 & 9010 & 10070 & 110$\pm$22 & 1106$\pm$91 & 685$\pm$37 & 1297.4$\pm$281 & (7.7$\pm$1.7)$\times$10$^{24}$ & (3.9$\pm$0.8)$\times$10$^{7}$ & (1.2$\pm$0.3)$\times$10$^{-9}$ \\
I16065--5158 & 3.98 & 5.2 & 6766 & 12338 & 150$\pm$43 & 225$\pm$17 & 85.8$\pm$4.9 & 124.9$\pm$37 & (5.0$\pm$1.5)$\times$10$^{23}$ & (2.0$\pm$0.6)$\times$10$^{6}$ & (9.5$\pm$2.8)$\times$10$^{-8}$ \\
I16071--5142 & 5.30 & 4.5 & 9010 & 22260 & 150$\pm$20 & 174$\pm$11 & 64.3$\pm$3.1 & 148.8$\pm$22 & (1.8$\pm$0.3)$\times$10$^{23}$ & (4.1$\pm$0.6)$\times$10$^{5}$ & (7.7$\pm$1.3)$\times$10$^{-8}$ \\
I16076--5134 & 5.30 & 4.5 & 9010 & 22790 & 121$\pm$32 & 87$\pm$11 & 21.9$\pm$2.2 & 92.6$\pm$27 & (1.1$\pm$0.3)$\times$10$^{23}$ & (2.4$\pm$0.7)$\times$10$^{5}$ & (7.2$\pm$2.1)$\times$10$^{-8}$ \\
I16164--5046 & 3.57 & 5.4 & 6069 & 9996 & 157$\pm$19 & 702$\pm$58 & 214$\pm$14 & 311.5$\pm$46 & (1.9$\pm$0.3)$\times$10$^{24}$ & (9.5$\pm$1.4)$\times$10$^{6}$ & (4.7$\pm$0.7)$\times$10$^{-9}$ \\
I16172--5028 & 3.57 & 5.4 & 6069 & 6426 & 110$\pm$32 & 643$\pm$38 & 475$\pm$18 & 409.8$\pm$122 & (6.0$\pm$1.8)$\times$10$^{24}$ & (4.7$\pm$1.4)$\times$10$^{7}$ & $\cdots$ \\
I16348--4654 & 12.09 & 5.4 & 20553 & 9672 & 213$\pm$12 & 167.2$\pm$5.6 & 135.3$\pm$2.8 & 624.6$\pm$42 & (4.0$\pm$0.3)$\times$10$^{24}$ & (2.1$\pm$0.1)$\times$10$^{7}$ & (3.7$\pm$0.4)$\times$10$^{-8}$ \\
I17016--4124c1 & 1.37 & 7.0 & 2603 & 1233 & 143$\pm$14 & 155$\pm$23 & 85.4$\pm$8.6 & 15.3$\pm$3 & (6.1$\pm$1.1)$\times$10$^{24}$ & (2.5$\pm$0.4)$\times$10$^{8}$ & (4.1$\pm$0.8)$\times$10$^{-9}$ \\
I17175--3544 & 1.34 & 7.0 & 2278 & 4556 & 183$\pm$8 & 622$\pm$68 & 247$\pm$20 & 45.9$\pm$5 & (1.3$\pm$0.2)$\times$10$^{24}$ & (1.5$\pm$0.2)$\times$10$^{7}$ & (3.1$\pm$0.4)$\times$10$^{-8}$ \\
I17220--3609 & 8.01 & 1.3 & 13617 & 12816 & 173$\pm$22 & 255$\pm$40 & 75.2$\pm$9.3 & 227.0$\pm$46 & (8.4$\pm$1.7)$\times$10$^{23}$ & (3.3$\pm$0.7)$\times$10$^{6}$ & (2.2$\pm$0.5)$\times$10$^{-8}$ \\
I17441--2822 & 8.10 & 0.2 & 11340 & 10530 & 144$\pm$7 & 2150$\pm$100 & 1240$\pm$41 & 1891.8$\pm$146 & (1.0$\pm$0.1)$\times$10$^{25}$ & (4.9$\pm$0.4)$\times$10$^{7}$ & (3.7$\pm$0.3)$\times$10$^{-9}$ \\
I18032--2032c4 & 5.15 & 3.4 & 7725 & 10815 & 140$\pm$13 & $\cdots$ & $\cdots$ & $\cdots$ & $\cdots$ & $\cdots$ & $\cdots$ \\
I18469--0132 & 5.16 & 4.7 & 7740 & 4644 & 136$\pm$15 & 68.1$\pm$8.3 & 32.7$\pm$2.8 & 63.5$\pm$10 & (1.8$\pm$0.3)$\times$10$^{24}$ & (1.9$\pm$0.3)$\times$10$^{7}$ & (6.2$\pm$2.0)$\times$10$^{-9}$ \\  
I18507+0110 & 1.56 & 7.1 & 2340 & 2028 & 200$\pm$11 & 1490$\pm$120 & 682$\pm$40 & 138.8$\pm$14 & (2.0$\pm$0.2)$\times$10$^{25}$ & (5.0$\pm$0.5)$\times$10$^{8}$ & (3.7$\pm$0.4)$\times$10$^{-9}$ \\  
I19078+0901c1 & 11.11 & 7.6 & 16665 & 18887 & 140$\pm$22 & 784$\pm$109 & 376$\pm$37 & 5881.0$\pm$1235 & (1.0$\pm$0.2)$\times$10$^{25}$ & (2.6$\pm$0.6)$\times$10$^{7}$ & (6.9$\pm$1.5)$\times$10$^{-10}$ \\
I19095+0930 & 6.02 & 5.8 & 9030 & 3010 & 121$\pm$3 & 101.5$\pm$5.9 & 95.5$\pm$3.2 & 180.8$\pm$12 & (1.2$\pm$0.1)$\times$10$^{25}$ & (2.0$\pm$0.1)$\times$10$^{8}$ & $\cdots$ \\
\hline
\end{tabular}}
\end{table}

\begin{table}
\renewcommand{\thetable}{C.4}
\centering
\caption{Basic physical parameters of hot cores associated with UC H{\sc ii} regions  without HC$_3$N* emission.}
\label{tab:param4}
\resizebox{1.34\textwidth}{!}{
\begin{tabular}{lcccccccccc}
\hline\hline
Source & Distance & R$\rm_{GC}$ & Resolution & R$\rm_{core}$ & T$ \rm _{d}$ & S$ \rm _{\nu} ^{int}$ & S$ \rm _{\nu} ^{peak}$ & M$ \rm _{core}$ & $ \rm N_{H_2}$ & n(H$_2$) \\
 & (kpc) & (kpc) & (au) & (au) & (K) & (mJy) & (mJy beam$^{-1}$) & (M$_{\sun}$) & (cm$^{-2}$) & (cm$^{-3}$) \\
\hline
I16351--4722 & 3.02 & 5.7 & 5134 & 4530 & 150$\pm$31 & 174$\pm$15 & 31$\pm$2.4 & 61.4$\pm$14 & (1.8$\pm$0.4)$\times$10$^{24}$ &(2.0$\pm$0.4)$\times$10$^{7}$ \\
I17016--4124c2 & 1.37 & 7.0 & 2603 & 2466 & 165$\pm$32 & 134.3$\pm$3.8 & 79.7$\pm$1.5 & 11.5$\pm$2 & (1.1$\pm$0.2)$\times$10$^{24}$ &(2.3$\pm$0.5)$\times$10$^{7}$ \\
I18032--2032c1 & 5.15 & 3.4 & 7725 & 4635 & 204$\pm$17 & $\cdots$ & $\cdots$ & $\cdots$ & $\cdots$ & $\cdots$ \\
\hline
\end{tabular}}
\end{table}
\end{landscape}

\end{CJK*}
\end{document}